\documentclass[final,3p,times]{elsarticle}

\usepackage{amssymb}

\biboptions{authoryear,semicolon,round}
\usepackage[utf8]{inputenc}
\usepackage{amsmath}
\usepackage{graphicx}
\usepackage{mathrsfs}
\usepackage{csquotes}
\usepackage{xcolor}
\usepackage[singlespacing]{setspace}
\usepackage[%
    pdftex,%
    colorlinks=true,%
    hyperindex,%
    plainpages=false,%
    pagebackref=true,%
    bookmarksopen,%
    bookmarksnumbered %
      ]{hyperref}
\renewcommand{\vec}[1]{\boldsymbol{#1}}
\DeclareUnicodeCharacter{2212}{-}
\providecommand{\added}[1]{{#1}}

\journal{Journal of the Mechanics and Physics of Solids}

\begin{document}

\begin{frontmatter}

\title{On the geometric stability of an inorganic nanowire\\ and an organic ligand shell\tnoteref{copyright}}

\tnotetext[copyright]{Typeset article available at \url{https://doi.org/10.1016/j.jmps.2018.07.017}. This manuscript version is made available under the \href{https://creativecommons.org/licenses/by-nc-nd/4.0/}{CC-BY-NC-ND 4.0 license}.}

\author[label1,label2]{Simon Bettscheider}
\address[label1]{INM - Leibniz Institute for New Materials, Campus D2.2, 66123 Saarbrücken, Germany}
\address[label2]{Colloid and Interface Chemistry, Saarland University, Campus, 66123 Saarbrücken, Germany}

\author[label1,label2]{Tobias Kraus}

\author[label3]{Norman A. Fleck\corref{cor1}}
\address[label3]{Cambridge University Engineering Department, Trumpington Street, Cambridge CB2 1PZ, UK}
\ead{naf1@eng.cam.ac.uk}
\cortext[cor1]{Corresponding author}

\begin{abstract}
The break-up of a nanowire with an organic ligand shell into discrete droplets is analysed in terms of the Rayleigh-Plateau instability. Explicit account is taken of the effect of the organic ligand shell upon the energetics and kinetics of surface diffusion in the wire. Both an initial perturbation analysis and a full numerical analysis of the evolution in wire morphology are conducted, and the governing non-dimensional groups are identified. The perturbation analysis is remarkably accurate in obtaining the main features of the instability, including the pinch-off time and the resulting diameter of the droplets. It is conjectured that the surface energy of the wire and surrounding organic shell depends upon both the mean and deviatoric invariants of the curvature tensor. Such a behaviour allows for the possibility of a stable nanowire such that the Rayleigh-Plateau instability is not energetically favourable. A stability map illustrates this. Maps are also constructed for the final droplet size and pinch-off time as a function of two non-dimensional groups that characterise the energetics and kinetics of diffusion in the presence of the organic shell. These maps can guide future experimental activity on the stabilisation of nanowires by organic ligand shells.
\end{abstract}


\end{frontmatter}



\section{Introduction}

Wires are prototypical components in electrical circuits: a metal wire is the simplest way of connecting two points electrically. The microelectronics industry has successfully evolved the miniaturization of metal and semiconductor wires (\enquote{interconnects}) using \emph{subtractive} processes based on lithography. Thin films are deposited and selectively etched to give in-plane features of dimension down to 10\,nm in the latest semiconductor technologies \citep{ITRS2011}. However, thin wires are inherently unstable. This can be traced to the fact that a circular cylinder of finite length has a larger surface area than a sphere of equal volume. The surface energy associated with this surface area is the driving force for the Rayleigh-Plateau instability: a long circular cylinder evolves into an array of spheres \citep{Plateau1873,Rayleigh1878}. The kinetic mechanism for the instability is either bulk or surface diffusion. The relative importance of these two mechanisms depends upon the relative diffusion constants and upon the wire diameter: at sufficiently small scale and at sufficiently low temperature, surface diffusion dominates \citep{Frost1982}. The Rayleigh-Plateau instability has been observed experimentally for a wide range of materials including copper \citep{Toimil-Molares2004}, silver \citep{Bid2005}, platinum \citep{Zhao2006}, gold \citep{Karim2006}, tin \citep{Shin2007}, nickel \citep{Zhou2009}, and cobalt \citep{Huang2010}. Recently, the Rayleigh-Plateau instability has also been observed in silicon nanowires, \citet{Barwicz2012}. These authors argue that the surface self-diffusivity of silicon is significantly increased by the presence of a reduced hydrogen environment, which provides a kinetic path for surface diffusion at temperatures exceeding 700\;$^{\circ}$C. 

\begin{figure}[ht]
    \centering
    \includegraphics[width=0.5\textwidth]{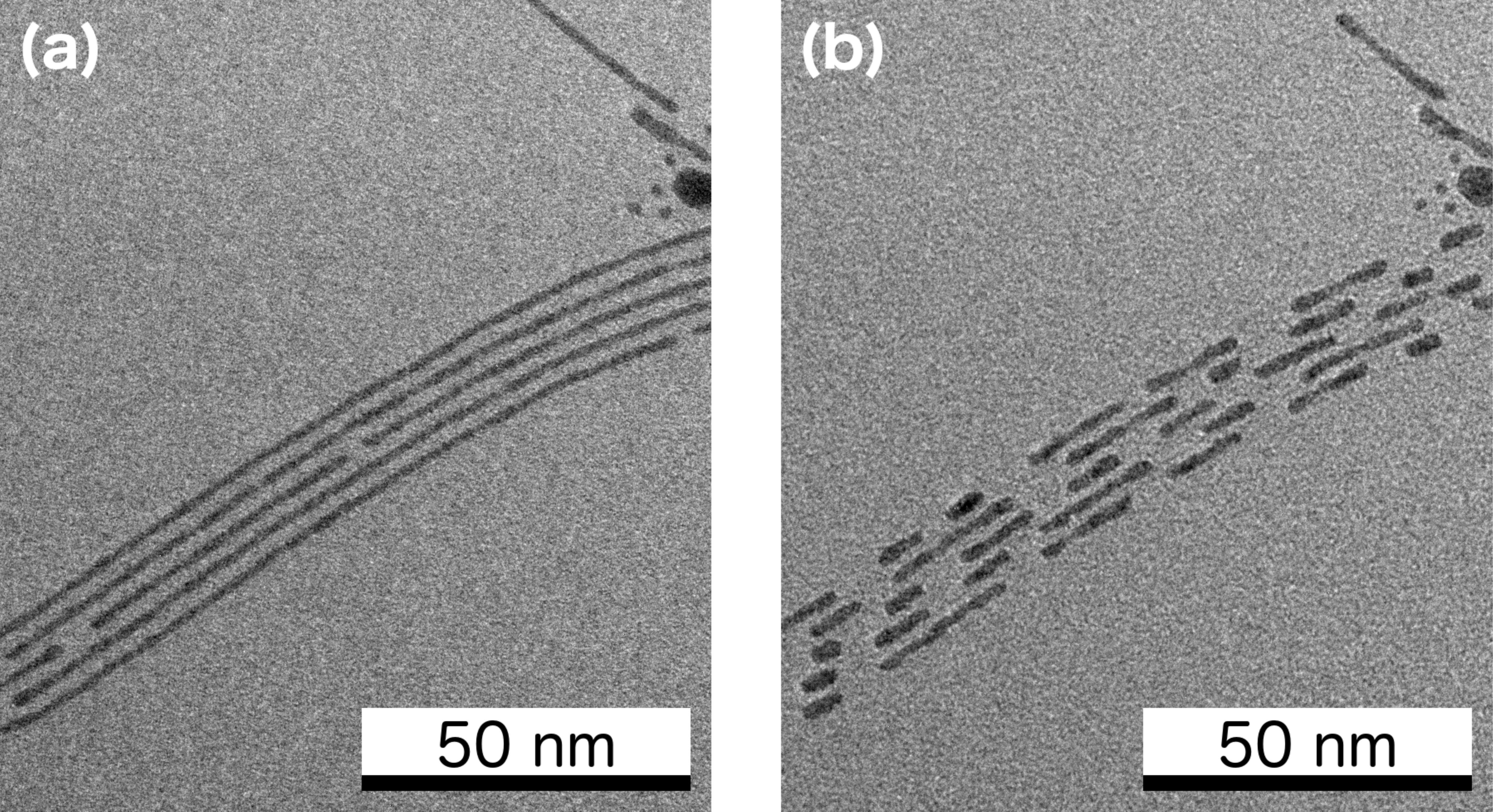}
    \caption{Transmission Electron Microscopy (TEM) micrographs of chemically synthesised gold nanowires on a carbon-coated copper grid. The synthesis followed that of \citet{Reiser2016,Reiser2017}, by adaption of the original protocol of \citet{Feng2009}. (a) Initial state, and (b) after electron beam heating for 212\;s at 200\;kV.} 
    \label{fig:TEMmicrograph}
\end{figure}

Recently, a new class of nanowires of diameter below 10\;nm has been made by chemical \emph{synthesis}. Such \enquote{ultrathin} nanowires form when solutions of metal or semiconductor salts are reduced in the presence of certain organic molecules. The detailed formation mechanism of the wires is a topic of present debate \citep{Cademartiri2009,Repko2012}; proposed routes involve the self-assembly of the organic molecules into micelle-like structures that template wire growth. Regardless of the precise mechanism of formation, the result is a metal core surrounded by a \enquote{ligand shell} of organic molecules. A common choice of organic ligand is oleylamine, a simple hydrocarbon chain with 18 carbon atoms and a single double bond at its centre, which forms an organic ligand shell of thickness approximately 2\;nm. To date, ultrathin nanowires made from gold \citep{Lu2008,Wang2008} (figure \ref{fig:TEMmicrograph}a), iron-platinum \citep{Wang2007}, silver \citep{Li2015}, calcium phosphate \citep{Sadasivan2005}, barium sulfate \citep{Hopwood1997}, tellurium \citep{Xi2006}, copper sulfide \citep{Liu2005}, bismuth sulfide \citep{Cademartiri2008}, antimony trisulfide \citep{Malakooti2008}, samarium oxide \citep{Yu2006}, and ruthenium \citep{Zhao2016} have all been synthesised in this manner. Such wires hold promise for new electronic devices such as mechanically flexible, optically transparent, or printable electronics \citep{Wang2008,PazosPerez2008,SanchezIglesias2012,Chen2013,Gong2014,Maurer2015,Maurer2016}. 

Ultrathin nanowires are prone to the Rayleigh-Plateau instability. Since the characteristic pinch-off time for the Rayleigh-Plateau instability scales with the fourth power of the wire radius \citep{Nichols1965,Nichols1965AIME}, ultrathin nanowires break-up into spheres at much shorter times and at lower temperatures than thicker wires. For example, \citet{Takahata2016} report the break-up of ultrathin gold nanorods of diameter of 2\;nm (and length of 6\,nm) into spheres after 5400\;s at 80\;$^\circ$C in liquid chloroform; similar observations have been reported by other authors \citep{Lu2008,Xu2013,Lacroix2014,Xu2017}. Electron beam heating in a transmission electron microscope (TEM) leads to the break-up of gold nanowires into a string of nanorods within four minutes, see figure \ref{fig:TEMmicrograph}b. 

\citet{Ciuculescu2009}, \citet{Huber2012}, \citet{WuJianbo2015}, and \citet{Takahata2016} all suggest that the break-up of ultrathin nanowires is delayed by the presence of an organic ligand shell. However, the mechanism of stabilisation remains unclear: it can be energetic or kinetic in nature, or a combination of the two. This lack of understanding inhibits a systematic search for ligands that could enhance wire stability to a point where storage at room temperature does not limit their applicability. The purpose of the present paper is to provide a framework for understanding the stabilisation mechanisms and to generate guiding principles for the selection of suitable ligands.

\section{Governing field equations}

The evolution in shape of a circular wire into spherical droplets by surface diffusion has been analysed by \citet{Nichols1965,Nichols1965AIME}: they treated the bulk of the wire as rigid and considered diffusion along the bare surface of the wire. This surface diffusion was driven by a gradient in chemical potential associated with local surface curvature. Finite shape changes were included in the analysis such that a long cylindrical wire breaks-up into an array of droplets. 

\begin{figure}[ht]
    \centering
    \includegraphics[width=1.0\textwidth]{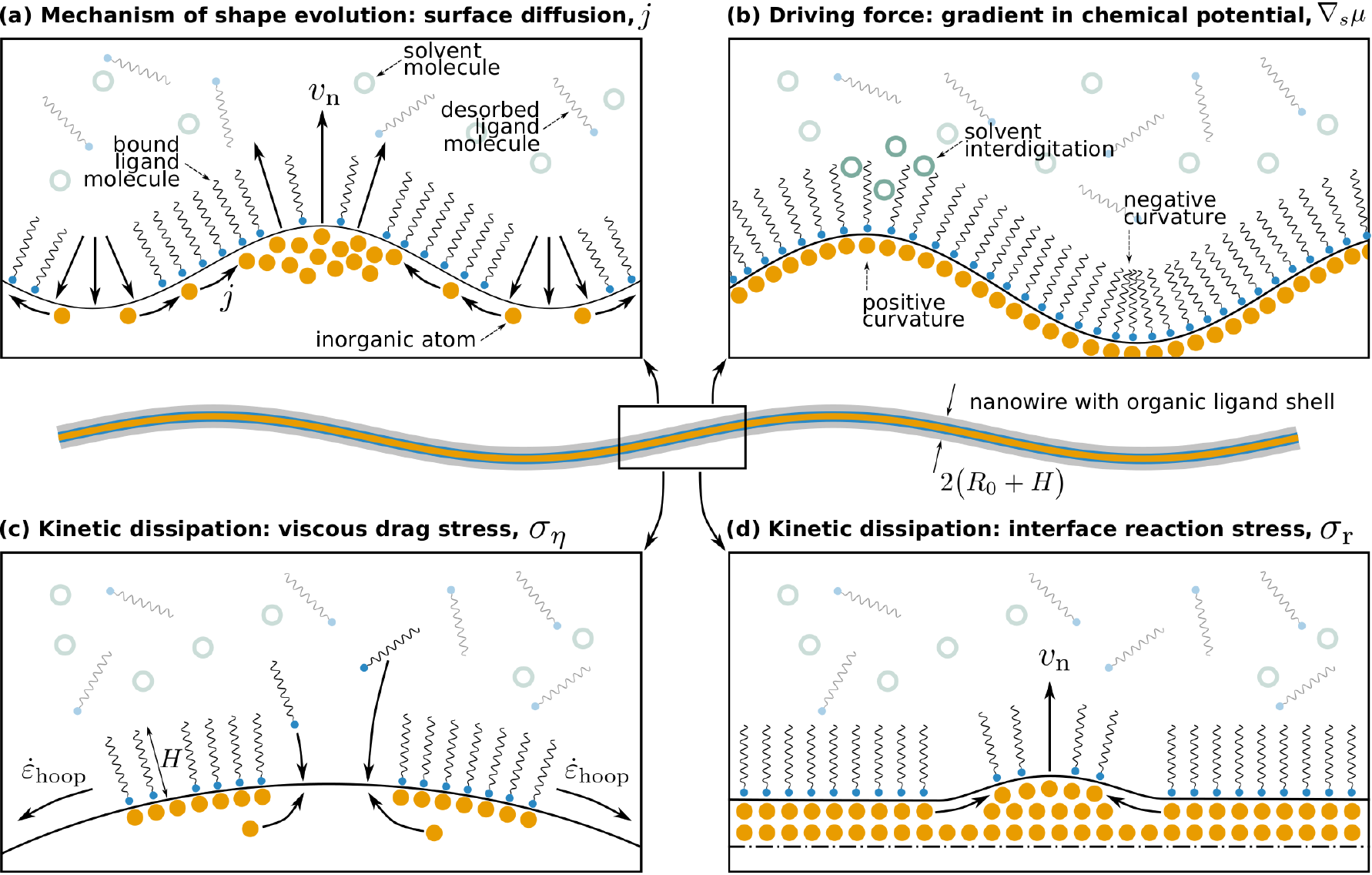}
    \caption{An inorganic nanowire and organic ligand shell in a solvent.  (a)  A flux $j$ of wire atoms diffuse along the wire surface and lead to an outward normal velocity $v_{\mathrm{n}}$. (b)  Diffusion is driven by the surface gradient in chemical potential $\mu$ (c) Adsorption and desorption of ligands is resisted by a viscous drag stress $\sigma_{\eta}$. (d) Plating of surface atoms is resisted by an interface reaction stress $\sigma_{\mathrm{r}}$.}
    \label{fig:physicalPicture}
\end{figure}

In the case of ultrathin nanowires, the wire surface has a more complex structure: the wire core is surrounded by the organic ligand shell, and the wire and shell are dispersed within an organic solvent, see figure \ref{fig:physicalPicture}. While the mechanism for shape evolution of the nanowire remains surface diffusion (see figure \ref{fig:physicalPicture}a), driven by a gradient in chemical potential, the presence of the shell introduces additional physical phenomena that need to be incorporated into the model. For example, the local packing arrangement of ligands within the shell, and thereby the free energy of the shell, depend upon local curvature (see figure \ref{fig:physicalPicture}b). In the present study, the surface energy $\gamma$ is treated as a function of surface curvature. The organic ligand shell is of fixed thickness $H$ but its circumference changes when the wire profile evolves. When the circumference of the shell changes, additional ligand molecules must assemble (or dissemble) into the shell and viscous losses occur, see figure \ref{fig:physicalPicture}c. This is modelled by a viscous drag stress $\sigma_{\eta}$ that depends upon the hoop strain rate of the organic ligand shell. Additionally, the plating of surface-diffusing atoms onto the surface of the wire involves an interface reaction and an attendant viscous drag, see figure \ref{fig:physicalPicture}d. This is idealised as a dissipative interface reaction, involving an interface reaction stress $\sigma_{\mathrm{r}}$ and its work conjugate, the normal velocity of the interface $v_{\mathrm{n}}$. These notions build upon previous models for grain growth due to surface diffusion, see for example \citet{Ashby1969}, \citet{Cocks1992}, and \citet{Cocks1998}.

In order to derive the governing equations for shape evolution of a nanowire surrounded by an organic ligand shell, we first introduce the geometry and kinematics of shape evolution of a small wire. Second, the chemical potential, interface reaction stress, and viscous drag stress are described and a governing ordinary differential equation (ODE) is developed for the outward normal velocity $v_{\mathrm{n}}$ along the surface of the wire in its current configuration. An updating scheme is then given for the wire profile as a function of time. 

\subsection{Geometry}

\begin{figure}[ht]
    \centering
    \includegraphics[width=1.0\textwidth]{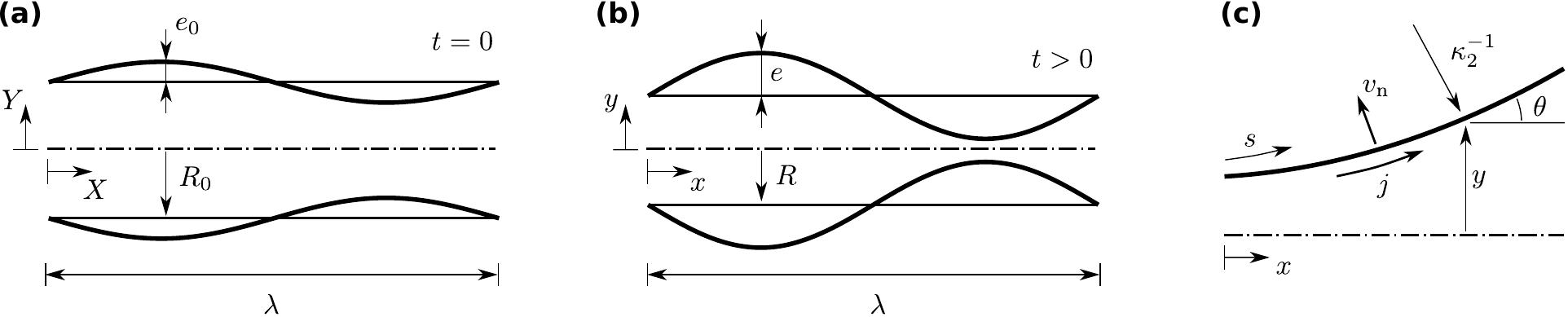}
    \caption{(a)  Nanowire of initial mean radius $R_0$ and sinusoidal perturbation of amplitude $e_0$ and wavelength $\lambda$, at time $t=0$. (b) In the current configuration at time $t>0$, the nanowire has a mean radius $R$ and a sinusoidal perturbation of amplitude $e$. (c) Geometry of current configuration.}
    \label{fig:geometry}
\end{figure}

Consider a circular cylindrical nanowire of radius $R_0$ and a surrounding organic ligand shell of thickness $H$, see figures \ref{fig:physicalPicture} and \ref{fig:geometry}a. Assume that the wire and shell maintain a circular cross-section when its shape evolves by the diffusion of atoms along the surface of the wire in the axial direction. Assume that the initial radius of the wire $Y$ depends upon the axial position $X$ according to
\begin{align}\label{eq:initialSinusoidalPert}
    Y = R_0 + e_0 \sin{(\omega X)}
\end{align}
in terms of an initial imperfection amplitude $e_0$ and wavenumber $\omega=2\pi/\lambda$, where $\lambda$ is the perturbation wavelength. It proves convenient to describe the initial shape of the wire in terms of intrinsic coordinates $(S,\Theta)$ rather than $(X,Y)$, where $S$ is the arc length along the surface profile and $\Theta$ is the inclination of the surface in the initial, reference configuration. The transformation is straightforward:
\begin{subequations}
\begin{align}
    \mathrm{d}S^2 &= \mathrm{d}X^2 + \mathrm{d}Y^2 \quad\text{and} \\
    \tan{\Theta} &= \frac{\mathrm{d}Y}{\mathrm{d}X}. 
\end{align}
\end{subequations}
such that
\begin{subequations}
\begin{align}
    \frac{\mathrm{d}X}{\mathrm{d}S} &= \cos{\Theta} \quad\text{and} \\
    \frac{\mathrm{d}Y}{\mathrm{d}S} &= \sin{\Theta}.
\end{align}
\end{subequations}

Now consider the current configuration at time $t$. Write $y(x)$ as the deformed shape at time $t$, in the current configuration. The challenge is to predict $y(x,t)$, due to surface diffusion. Write $s$ as the arc length along the wire surface and $\theta$ as the inclination of the surface of the wire, as shown in figure \ref{fig:geometry}c. We seek $s=s(S,t)$ and $\theta=\theta(S,t)$.

The two principal curvatures of the wire surface are
\begin{subequations}
\begin{align}
    \kappa_{1} &= \frac{\cos{\theta}}{y}, \quad\text{and} \label{eq:kap1}\\
    \kappa_{2} &= -\frac{\partial\theta}{\partial s}, \label{eq:kap2}
\end{align}
\end{subequations}
and the stretch factor $\Lambda$ relates $s$ in the current configuration to $S$ in the initial configuration, such that
\begin{align}\label{eq:stretchFactor}
    \Lambda = \frac{\partial s}{\partial S}.
\end{align}

\subsection{Nanowire shape evolution by surface diffusion}

The dominant mechanism for shape evolution of a nanowire in the presence of an organic ligand shell is taken to be surface diffusion, see for example \citet{Ciuculescu2009}. We begin by relating the surface gradient of the volumetric surface diffusive flux $\vec{j}$ to the outward normal velocity $v_{\mathrm{n}}$ of the surface of the wire as demanded by mass conservation, such that
\begin{equation} \label{eq:vel}
    \vec{\nabla}_{s} \cdot \vec{j} + v_{\mathrm{n}} = 0.
\end{equation}
Now, the force $\vec{f}$ that drives surface diffusion is the surface gradient of the chemical potential $\mu$,
\begin{align}\label{eq:driveForce}
    \vec{f} = \vec{\nabla}_{s}\mu
\end{align}
and, in the absence of other kinetic dissipation processes, the flux $\vec{j}$ scales with $\vec{f}$ according to
\begin{align}\label{eq:diffFluxDrive}
    \vec{j} = -\frac{\mathscr{D}\vec{f}}{\Omega}, 
\end{align}
in terms of a mobility $\mathscr{D}/\Omega$, where $\mathscr{D}$ is the interface diffusion constant in units of $\mathrm{m^6J^{-1}s^{-1}}$ and $\Omega$ is the atomic volume. $\mathscr{D}$ is related to the interface diffusion coefficient $D_{\mathrm{b}}$ and the effective thickness of the boundary $\delta_{\mathrm{b}}$ according to $\mathscr{D}=D_{\mathrm{b}}\delta_{\mathrm{b}}\Omega/(kT)$, where $k$ is the Boltzmann constant and $T$ is absolute temperature. We take $\Omega=\delta_{\mathrm{b}}^3$. 

Limit attention to a solid of revolution, and consequently equations \eqref{eq:vel} and \eqref{eq:driveForce} reduce to
\begin{subequations}
\begin{align}
    v_{\mathrm{n}} &= - \frac{1}{y}\frac{\partial(jy)}{\partial s},\quad\text{and} \label{eq:vnjCyl}\\
    f &= \frac{\partial\mu}{\partial s}, \label{eq:diffFluxCyl}
\end{align}
\end{subequations}
respectively, as discussed by \citet{Nichols1965}. 

Now, if additional dissipation processes are at play, $f$ not only drives the diffusive flux $j$ but also needs to drive the additional drag processes. Here, we shall assume the presence of an interface reaction associated with the outward normal velocity of the wire surface $v_{\mathrm{n}}$, and viscous drag associated with the hoop strain rate of the wire surface, so that \eqref{eq:diffFluxDrive} now reads
\begin{align}\label{eq:netDriveForce}
    j = -\frac{\mathscr{D}}{\Omega}\big(f - f_{\mathrm{r}} - f_{\eta}\big),
\end{align}
where $f_{\mathrm{r}}$ and $f_{\eta}$ are the forces needed to drive the interface reaction and the viscous dissipation, respectively. Following \citet{Ashby1969}, \citet{Cocks1992}, and \citet{Cocks1998}, the latter two forces can be written as
\begin{subequations}
\begin{align}
    f_{\mathrm{r}} &= -\Omega \frac{\partial\sigma_{\mathrm{r}}}{\partial s}, \quad\text{and} \label{eq:fSigma}\\ 
    f_{\eta} &= -\Omega \frac{\partial\sigma_{\eta}}{\partial s}, \label{eq:fEta}
\end{align}
\end{subequations}
where $\sigma_{\mathrm{r}}$ and $\sigma_{\eta}$ are the stresses normal to the interface that drive the interface reaction and the viscous drag, respectively. Insert equations \eqref{eq:fSigma} and \eqref{eq:fEta} into equation \eqref{eq:netDriveForce} to give
\begin{align}\label{eq:jEnerDiss}
    j = -\frac{\mathscr{D}}{\Omega} \frac{\partial\mu}{\partial s} 
    - \mathscr{D}\frac{\partial\sigma_{\mathrm{r}}}{\partial s} 
    - \mathscr{D}\frac{\partial\sigma_{\eta}}{\partial s}.
\end{align}
The remainder of this section will deal with the determination of $\mu$, $\sigma_{\mathrm{r}}$, and $\sigma_{\eta}$. 

\subsection{The chemical potential $\mu$}

The wire is coated by an organic ligand shell and we assume that a representative species of the wire-shell-interface comprises a surface metallic atom and a ligand molecule bound to it. The chemical potential $\mu$ is, by definition, the energy required to bring a metallic atom from the bulk to the surface of the wire and to bring its partner organic ligand molecule from remote solvent to the bound state of the metal-organic complex. 

Assume that the free energy of the organic ligand molecule depends upon its local configuration, as characterised by the curvature $(\kappa_1,\kappa_2)$ of the wire at this location. We envisage that the free energy of the organic ligand molecule is less when bound to a sphere of total curvature $2\kappa_{\mathrm{s}}$ (such that $\kappa_1=\kappa_2=\kappa_{\mathrm{s}}>0$) than for a cylinder of total curvature $\kappa_{\mathrm{s}}$ (such that $\kappa_1=\kappa_{\mathrm{s}}$ and $\kappa_2 = 0$). Also, the free energy of a surface of saddle point shape with zero total curvature (such that $\kappa_1=-\kappa_2=\kappa_{\mathrm{s}}$) is higher still. An empirical relation for the chemical potential of the metal-organic-ligand complex that captures these features is
\begin{equation}\label{eq:muGamma}
    \mu = 2\Omega\gamma\kappa_{\mathrm{m}},
\end{equation}
where $\kappa_{\mathrm{m}}=(\kappa_1+\kappa_2)/2$ is the mean curvature, and the surface energy $\gamma$ depends upon an effective curvature $\kappa_{\mathrm{e}}$. The effective curvature is taken to be a function of the two invariants of the curvature tensor and we shall assume that it adopts the simple quadratic form
\begin{align}
    \kappa_{\mathrm{e}}^2 &= \alpha\kappa_{\mathrm{m}}^2 + (1-\alpha)\kappa_{\mathrm{d}}^2, \label{eq:kapE}
\end{align}
where the deviatoric curvature reads $\kappa_{\mathrm{d}}=(\kappa_1-\kappa_2)/2$ and the  \enquote{shape factor} $\alpha$ is between zero and unity. Note that the choice $\alpha=1$ implies that $\kappa_{\mathrm{e}}=\lvert\kappa_{\mathrm{m}}\rvert$ while $\alpha=0$ implies $\kappa_{\mathrm{e}}=\lvert\kappa_{\mathrm{d}}\rvert$. 

It remains to stipulate a functional form for $\gamma(\kappa_{\mathrm{e}})$. We adopt the empirical choice
\begin{align}\label{eq:gammaFunctional}
    \gamma(\kappa_{\mathrm{e}}) = \gamma_0
    \exp{\bigg(c\Big(\frac{\kappa_{\mathrm{e}}}{\kappa_0} - 1 \Big) \bigg)},
\end{align}
where $\gamma_0$, $c$, and $\kappa_0$ are material constants. The sensitivity of $\gamma$ to $\kappa_{\mathrm{e}}$ is largely captured by the value of $c$ and we shall assume both positive and negative values for $c$ in our analysis below. 

\added{The authors recognise that the continuum approach loses validity as the length scales approach atomistic dimensions. However, experimental evidence on the transition from cylindrical wires to spheres suggests that the wires behave as a continuum, with a smooth surface that can be described by continuous functions, such as that used in \eqref{eq:kapE}. The present study could be complemented by molecular dynamics (MD) simulations, but it will be a challenge to match the timescales of MD simulations (on the order of picoseconds) to the timescales as observed in an experiment (on the order of 100-1000 seconds). The current approach is phenomenological, and it may be possible to calibrate the parameters of the model using predictions from MD simulations of the wires' interfaces.}

\subsection{The interface reaction stress $\sigma_{\mathrm{r}}$ and viscous stress $\sigma_{\eta}$}

\added{The plating/removal of atoms onto/from a surface involves local rearrangement of atoms as they change configuration from a bulk co-ordination number on the interior to a reduced co-ordination number on the surface. Such rearrangements involve viscous drag at low temperatures, modelled by an interface reaction stress.} Following \citet{Cocks1992} and \citet{Cocks1998}, the reaction stress is assumed to take the form of a power law according to
\begin{equation}\label{eq:sigmaR}
    \sigma_{\mathrm{r}} = \sigma_{\mathrm{r0}}\bigg(\frac{v_{\mathrm{n}}}{\lvert v_{\mathrm{n}}\rvert} \bigg) \bigg(\frac{\lvert v_\mathrm{n}\rvert}{v_{\mathrm{r0}}}\bigg)^{M},
\end{equation}
in terms of a reference velocity $v_{\mathrm{r0}}$, a reaction strength $\sigma_{\mathrm{r0}}$, and an exponent $M$. \added{An appropriate choice for the exponent is $M=1$ \citep{Ashby1969,Cocks1992,Cocks1998}; it leads to considerable simplification of the governing equations for shape evolution of the nanowire.}

In order to obtain the viscous drag stress $\sigma_{\eta}$ we consider an organic ligand shell subjected to equi-biaxial tension. A derivation of $\sigma_{\eta}$ as a function of $v_{\mathrm{n}}$ is given in the appendix. Here, we simply state the results. The viscous drag stress reads 
\begin{equation}\label{eq:sigmaEta}
    \sigma_{\eta} = \frac{\partial\phi_{\eta}}{\partial v_{\mathrm{n}}} = 4 H \eta \kappa_{\mathrm{m}}^2 v_{\mathrm{n}}, 
\end{equation}
where the viscosity $\eta$ is taken to be a function of curvature, 
\begin{align}\label{eq:etaFunctional}
    \eta(\kappa_{\mathrm{e}}) = \eta_0
    \exp{\bigg(d\Big(\frac{\kappa_{\mathrm{e}}}{\kappa_0} - 1 \Big) \bigg)}
\end{align}
in terms of the material constants $\eta_0$, $d$ and $\kappa_0$.

\subsection{Governing equation for normal velocity}

We proceed to insert equations \eqref{eq:muGamma}, \eqref{eq:sigmaR}, and \eqref{eq:sigmaEta} into equations \eqref{eq:jEnerDiss} and \eqref{eq:vnjCyl} to obtain a non-linear second order ODE in $v_{\mathrm{n}}$,
\begin{align}\label{eq:govEquVN}
    \frac{\Lambda y v_{\mathrm{n}} }{\mathscr{D}}
    - 
    \frac{\partial}{\partial S} 
    \Bigg( 
    \frac{y}{\Lambda}\frac{\partial}{\partial S} 
    \Big[ 
    \sigma_{\mathrm{r0}} \bigg(\frac{v_{\mathrm{n}}}{\lvert v_{\mathrm{n}} \rvert} \bigg)\bigg(\frac{\lvert v_\mathrm{n}\rvert}{v_{\mathrm{r0}}}\bigg)^{M}
    +
    4 H \eta \kappa_{\mathrm{m}}^2 v_{\mathrm{n}}
    +
    2\gamma\kappa_{\mathrm{m}}
    \Big] 
    \Bigg)
    =  
    0.
\end{align}

\section{Perturbation analysis}

In order to identify the dominant non-dimensional groups and assess their relative importance in dictating wire stability, a linear perturbation analysis is now conducted. Later, the predictions of the perturbation analysis will be compared with a full numerical solution. Consider a cylindrical wire of initial shape as given by equation \eqref{eq:initialSinusoidalPert}, and as depicted in figure \ref{fig:geometry}a. Assume that the perturbation evolves with time $t$ as illustrated in figure \ref{fig:geometry}b such that
\begin{align}\label{eq:yPert}
    y(x,t) = R(t) + e(t)\sin{(\omega x)},
\end{align}
where $x=X$, and $R(t)$ and $e(t)$ are to be determined. Our aim is to re-express the ODE \eqref{eq:govEquVN} in $v_{\mathrm{n}}$ as an ODE in $\mathrm{d}e/\mathrm{d}t$. Conservation of mass dictates that
\begin{equation}
   \int_{0}^{\lambda} Y^2 \mathrm{d}X 
   = 
   \int_{0}^{\lambda} y^2(t) \mathrm{d}x
\end{equation}
and consequently
\begin{align}\label{eq:R(e)Cahn}
    R^2 + \frac{1}{2}e^2 = R_0^2 + \frac{1}{2}e_0^2.
\end{align}
The two principal curvatures, as defined in equations \eqref{eq:kap1} and \eqref{eq:kap2}, evolve with time according to
\begin{subequations}
\begin{align}\
    \kappa_{1} &= \Bigg(1 + (\omega e)^2 \cos^2{(\omega x)} \Bigg)^{-1/2}\Bigg(\bigg(R_0^2 + \frac{1}{2}e_0^2 - \frac{1}{2}e^2\bigg)^{\frac{1}{2}} + e\sin{(\omega x)} \Bigg)^{-1} \quad\text{and} \label{eq:kap1pert}\\
    \kappa_{2} &= \omega^2 e \sin{(\omega x)}\Big(1 + (\omega e)^2 \cos^2{(\omega x)} \Big)^{-3/2}, \label{eq:kap2pert}
\end{align}
\end{subequations}
upon making use of the identity $\partial y/\partial x=\tan{\theta}$ and the relations \eqref{eq:yPert} and \eqref{eq:R(e)Cahn}. Now expand $\kappa_1$ and $\kappa_2$ in terms of increasing powers of $e$ and neglect terms of order $e^2$ and higher to obtain 
\begin{subequations}
\begin{align}
    R_0\kappa_1 &= 1 - \frac{e}{R_0}\sin{(\omega x)} + \mathcal{O}(e^2), \quad\text{and} \label{eq:kap1expand} \\
    R_0\kappa_{2} &= \frac{e}{R_0} (R_0\omega)^2 \sin{(\omega x)} + \mathcal{O}(e^3), \label{eq:kap2expand}
\end{align}
\end{subequations}
so that the mean curvature reads 
\begin{align}
    2R_0\kappa_{\mathrm{m}} &= 1 + \frac{e}{R_0}\bigg( (R_0\omega)^2 - 1 \bigg)\sin{(\omega x)} + \mathcal{O}(e^2), 
\end{align}
and the effective curvature reduces to
\begin{align}
    2R_0\kappa_{\mathrm{e}} &= 1
    + \frac{e}{R_0}\Bigg((2\alpha-1)(R_0\omega)^2 - 1 \Bigg)\sin{(\omega x)} + \mathcal{O}(e^2) . 
\end{align}
Likewise, expand the surface energy $\gamma(\kappa_{\mathrm{e}})$ and the viscosity $\eta(\kappa_{\mathrm{e}})$ as defined in \eqref{eq:gammaFunctional} and \eqref{eq:etaFunctional}, respectively, to give
\begin{subequations}
\begin{align}
    \gamma(\kappa_{\mathrm{e}}) &= \gamma_0 + \frac{\gamma_0 c e}{R_0} \Big((2\alpha-1)(R_0\omega)^2 - 1 \Big)\sin{(\omega x)} + \mathcal{O}(e^2), \label{eq:gammaExpanded} \quad\text{and} \\
    \eta(\kappa_{\mathrm{e}}) &= \eta_0 + \frac{\eta_0 d e}{R_0} \Big((2\alpha-1)(R_0\omega)^2 - 1 \Big)\sin{(\omega x)} + \mathcal{O}(e^2), 
\end{align}
\end{subequations}
upon taking $\kappa_0=1/(2R_0)$. 

Now, for small perturbations, we may write 
\begin{align}
    \frac{1}{y}\frac{\partial}{\partial s} \Bigg( y\frac{\partial}{\partial s} \Bigg) \approx \frac{\partial^2}{\partial x^2}, 
\end{align} 
so that, by limiting the analysis to cases where $M=1$ in equation \eqref{eq:sigmaR}, the governing ODE \eqref{eq:govEquVN} becomes
\begin{align}\label{eq:govEqPert}
    \frac{1}{\mathscr{D}} v_{\mathrm{n}} 
    - \frac{\partial^2}{\partial x^2} \Big[ \frac{\sigma_{\mathrm{r0}}}{v_{\mathrm{r0}}} v_\mathrm{n} \Big] 
    - \frac{\partial^2}{\partial x^2} \Big[ 4 H \eta \kappa_{\mathrm{m}}^2 v_{\mathrm{n}} \Big] 
    =  \frac{\partial^2}{\partial x^2} \Big[ 2\gamma\kappa_{\mathrm{m}} \Big] .
\end{align}
The outward normal velocity $v_{\mathrm{n}}$ scales with $\dot{R}$ and $\dot{e}$ according to
\begin{align}\label{eq:vnLinearized}
    v_{\mathrm{n}} = \dot{R} + \dot{e}\sin{(\omega x)}, 
\end{align}
where $\dot{(\;)}$ denotes $\mathrm{d}(\;)/\mathrm{d}t$. Recall that $\dot{R}$ is a function of $\dot{e}$ as dictated by \eqref{eq:R(e)Cahn} such that
\begin{align}\label{eq:RDotCahn}
    \dot{R} = -\frac{1}{2}\frac{e}{R_0}\dot{e} + \mathcal{O}(e^2).
\end{align}
Since we neglect terms of order $e^2$ and higher, we can neglect the contribution from $\dot{R}$ to $v_{\mathrm{n}}$, and \eqref{eq:govEqPert} can be linearised to 
\begin{align}\label{eq:govEquPert}
    \frac{R_0^4}{\mathscr{D}\gamma_0} \frac{\dot{e}}{R_0} 
    \bigg(
    1 
    + \frac{\mathscr{D}\sigma_{\mathrm{r0}}(R_0\omega)^2}{v_{\mathrm{r0}}R_0^2} 
    + \frac{\mathscr{D} H \eta_0(R_0\omega)^2}{R_0^4} 
    \bigg)
    =
    \frac{e(R_0\omega)^2}{R_0}
    \bigg(
    \Big[1 + (2\alpha-1)c \Big](R_0\omega)^2 - 1 - c\Big]
    \bigg),
\end{align}
upon making use of \eqref{eq:vnLinearized} to \eqref{eq:gammaExpanded}. 

We proceed to introduce the non-dimensional geometric variables 
\begin{align}\label{eq:nonDimGeom}
    \bar{\omega} = R_0\omega,
    \quad
    \bar{e} = e/R_0,
\end{align}
the non-dimensional time $\bar{t}$ as
\begin{align}\label{eq:nonDimTime}
    \bar{t} = (1+c)\frac{\mathscr{D}\gamma_0 t}{R_0^4}, 
\end{align}
and the non-dimensional material groups
\begin{align}\label{eq:nonDimMat}
    \chi_1 = \frac{1+c}{1+(2\alpha-1)c}, 
    \quad
    \chi_2 = \bar{\sigma}_{\mathrm{r0}} + \bar{\eta}_0,
    \quad
    \bar{\sigma}_{\mathrm{r}0} = \frac{\mathscr{D}\sigma_{\mathrm{r}0}}{R_0^2 v_{\mathrm{r}0}},
    \quad\text{and}\quad
    \bar{\eta}_0 = \frac{\mathscr{D} H \eta_0}{R_0^4}.
\end{align}
Note that $\chi_1$ contains only material parameters associated with the energetic parameter $c$ and the shape of the $\kappa_{\mathrm{e}}(\kappa_{\mathrm{m}},\kappa_{\mathrm{d}})$ locus. In contrast, $\chi_2$ includes only the kinetic terms $\bar{\sigma}_{\mathrm{r0}}$ and $\bar{\eta}_0$. Equation \eqref{eq:govEquPert} can be re-written to express the perturbation growth rate for a given perturbation wavenumber in the compact form 
\begin{align}\label{eq:eDotOvere}
    \frac{1}{\bar{e}}\frac{\mathrm{d}\bar{e}}{\mathrm{d}\bar{t}} = \frac{\bar{\omega}^2 - \chi_1^{-1}\bar{\omega}^4}
    {1 + \chi_2 \bar{\omega}^2 }.
\end{align}

\begin{figure}[ht]
    \centering
    \includegraphics[width=0.8\textwidth]{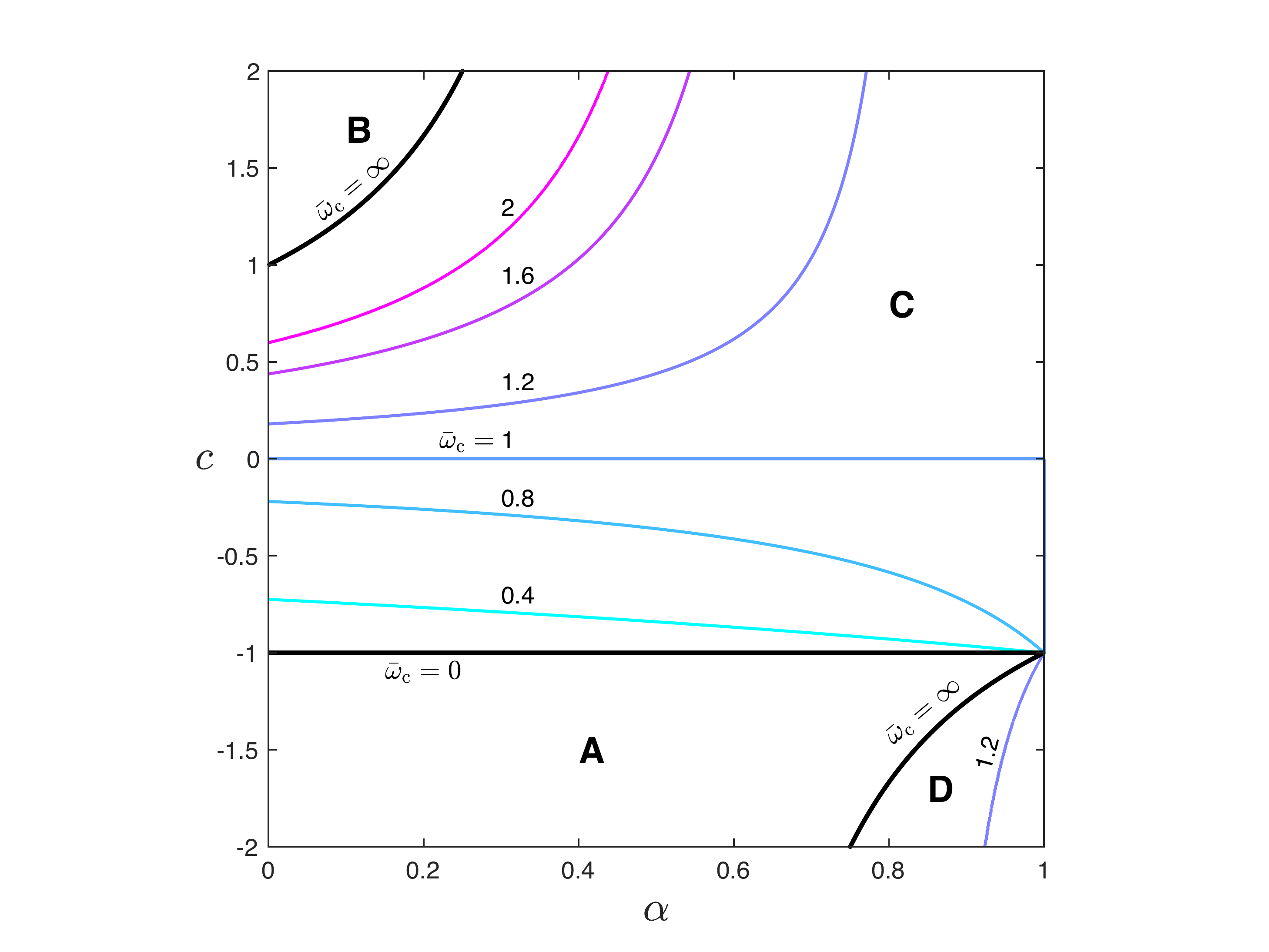}
    \caption{The four regimes of stability, with contours of critical wavenumber $\bar{\omega}_{\mathrm{c}}$.  Regime A: stable such that $\mathrm{d}\bar{e}/\mathrm{d}\bar{t}<0$ for all $\bar{\omega}$. Regime B: unstable such that $\mathrm{d}\bar{e}/\mathrm{d}\bar{t}>0$ for all $\bar{\omega}$.  Regime C: stable for $\bar{\omega}>\bar{\omega}_{\mathrm{c}}$.  Regime D: stable for $\bar{\omega}<\bar{\omega}_{\mathrm{c}}$.}
    \label{fig:analyticalRegimeMap}
\end{figure}

We proceed to evaluate the stability of a perturbation for any assumed wavenumber $\bar{\omega}$ by evaluating the sign of the perturbation growth rate $\mathrm{d}\bar{e}/\mathrm{d}\bar{t}$. A positive value of $\mathrm{d}\bar{e}/\mathrm{d}\bar{t}$ implies instability of the wire profile, whereas a negative value implies stability. Upon examining \eqref{eq:eDotOvere}, we find that the wire is stable for all $\bar{\omega}$ provided $(\alpha,c)$ satisfy the values
\begin{subequations}
\begin{align}
    0 < \alpha < \frac{1}{2} &\quad\text{and}\quad c < -1, \quad\text{or} \\
    \frac{1}{2} < \alpha < 1 &\quad\text{and}\quad \frac{-1}{2\alpha-1} < c < -1,
\end{align}
\end{subequations}
and we label this regime A in $(\alpha,c)$ space. The wire is unstable for all $\bar{\omega}$, when
\begin{align}
    0 < \alpha < \frac{1}{2} \quad\text{and}\quad c > \frac{-1}{2\alpha-1},
\end{align}
and this is labelled regime B in $(\alpha,c)$ space. The critical perturbation wavenumber $\bar{\omega}_{\mathrm{c}}$ at which $\mathrm{d}\bar{e}/\mathrm{d}\bar{t}=0$ follows immediately from \eqref{eq:eDotOvere} as
\begin{align}\label{eq:omgCrit}
    \bar{\omega}_{\mathrm{c}} = \big(\chi_1\big)^{1/2}.
\end{align}
Note that the perturbation is conditionally stable for wavenumbers $\bar{\omega}>\bar{\omega}_{\mathrm{c}}$ for $(\alpha,c)$ in regime C of $(\alpha,c)$ space:
\begin{subequations}
\begin{align}
    0 < \alpha < \frac{1}{2} &\quad\text{and}\quad c < \frac{-1}{2\alpha-1}, \quad\text{or} \\
    \frac{1}{2} < \alpha < 1 &\quad\text{and}\quad c > \frac{-1}{2\alpha-1}. 
\end{align}
\end{subequations}
Finally, the wire is conditionally stable for wavenumbers $\bar{\omega}<\bar{\omega}_{\mathrm{c}}$ in regime D, where
\begin{subequations}
\begin{align}
    0 < \alpha < \frac{1}{2} &\quad\text{and}\quad c > \frac{-1}{2\alpha-1}, \quad\text{or} \\
    \frac{1}{2} < \alpha < 1 &\quad\text{and}\quad c < \frac{-1}{2\alpha-1}.
\end{align}
\end{subequations}
The regimes A to D are marked on $(\alpha,c)$ space in figure \ref{fig:analyticalRegimeMap} and contours of the critical perturbation wavenumber $\bar{\omega}_{\mathrm{c}}$ are included. 

Salient features of the map are now discussed. For the choice $c=0$, the surface energy $\gamma$ is constant and $\bar{\omega}_{\mathrm{c}}=1$, as predicted by \citet{Nichols1965AIME}. The range of wavenumbers $\bar{\omega}>\bar{\omega}_{\mathrm{c}}$, for which the wire is stable, shrinks as $c$ increases, such that for sufficiently large $c$ (and small $\alpha$) regime B is entered and the wire is unconditionally unstable. Recall that a positive value of $c$ leads to an increase in $\gamma$ with increasing effective curvature $\kappa_{\mathrm{e}}$, see \eqref{eq:gammaFunctional}. Alternatively, a negative value of $c$ implies a reduction in $\gamma$ with increasing $\kappa_{\mathrm{e}}$, and this choice of value for $c$ stabilises the wire profile. Thus, for $c<-1$, regime A is entered and the wire is unconditionally stable for all $\bar{\omega}$. Regime D is small in extent: this regime of conditional stability for $\bar{\omega}<\bar{\omega}_{\mathrm{c}}$ exists only for large negative $c$ and large positive $\alpha$. 

\added{It is emphasised that the above perturbation analysis is based on \eqref{eq:govEquVN}, which contains both kinetic and energetic terms. Simplification of \eqref{eq:govEquVN} leads to the first order equation \eqref{eq:eDotOvere}, and it is noted that the numerator on the r.h.s. of \eqref{eq:eDotOvere} involves only energetic terms while the denominator has only kinetic terms. Consequently, the stability regimes, as plotted in figure \ref{fig:analyticalRegimeMap}, depend only on the energetic terms, and are independent of the various assumptions made in the kinetic part of the model. Thus, the response plotted in figure \ref{fig:analyticalRegimeMap} has broad applicability and the conclusions drawn from the perturbation analysis are rather general.}

\subsection{Prediction of final droplet size and pinch-off time}

The above perturbation analysis can be used to estimate the radius $R_{\mathrm{s}}$ of the spherical droplet that arises from break-up of the wire from the fastest growing perturbation. Also, an estimate for the pinch-off time $\bar{t}_{\mathrm{p}}$ required to convert the wire into an array of such spherical droplets can be obtained. 

\begin{figure}[ht]
    \centering
    \includegraphics[width=1.0\textwidth]{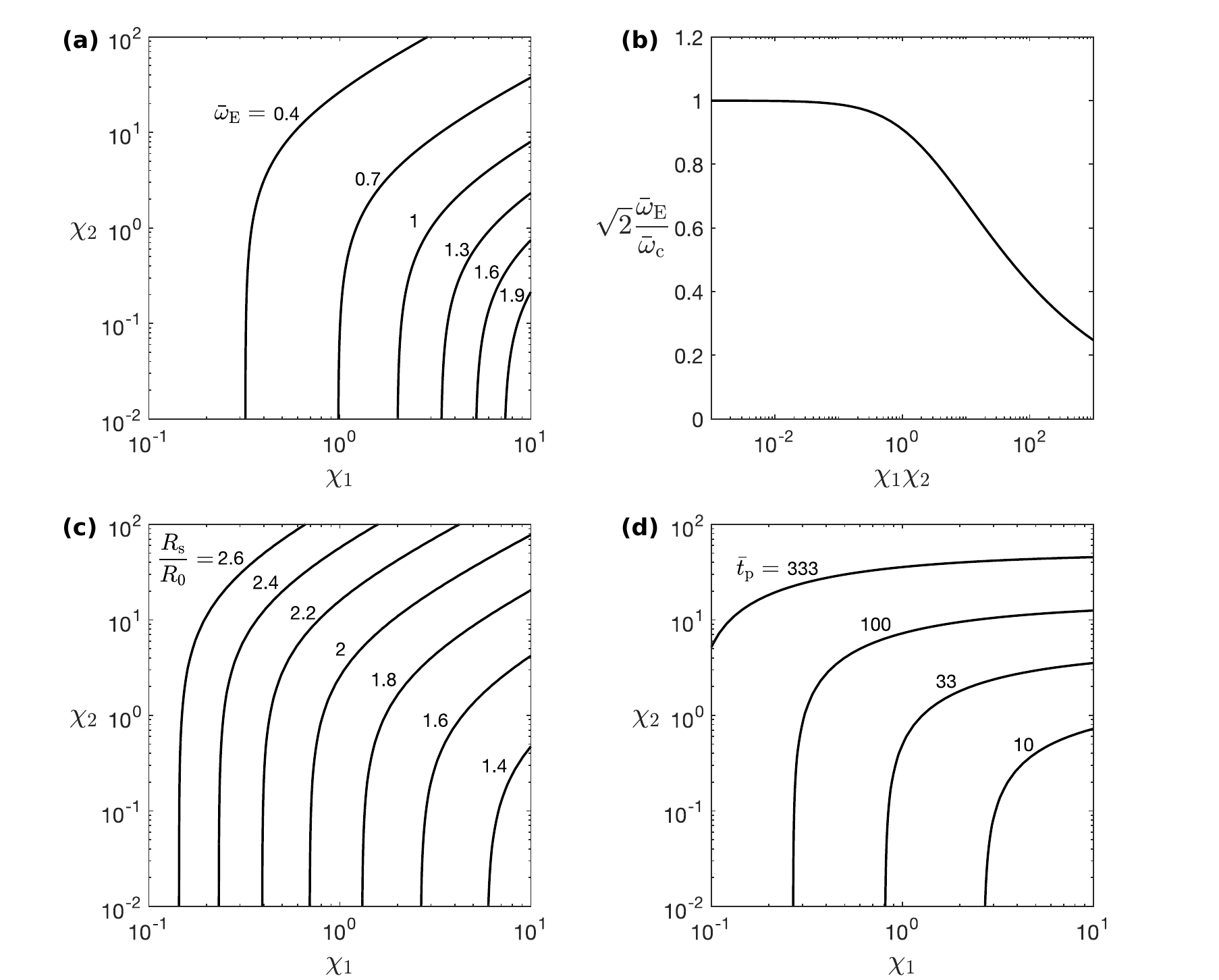}
    \caption{(a)  Contours of the fastest growing wavenumber $\bar{\omega}_{\mathrm{E}}$ in the $(\chi_1\chi_2)$ plane.  (b)  $\bar{\omega}_{\mathrm{E}}/\bar{\omega}_{\mathrm{c}}$ versus $\chi_1 \chi_2$.  (c)  Contours of radius of spherical droplet $R_{\mathrm{s}}$, normalised by the initial wire radius $R_0$, in the $(\chi_1\chi_2)$ plane.  (d)  Contours of pinch-off time $\bar{t}_{\mathrm{p}}$ in the $(\chi_1\chi_2)$ plane.}
    \label{fig:analyticalPerturbation}
\end{figure}

It is recognised that a spherical droplet is attained only at $t \rightarrow \infty$. However, we can identify a finite pinch-off time $\bar{t}_{\mathrm{p}}$ by assuming arbitrarily that the wire pinches off when $\bar{y}$ at any location along the wire has dropped to a selected value of $(1-\bar{e}_{\mathrm{p}})$. The radius $R_{\mathrm{s}}$ of the final spherical droplet is set by the value of wavenumber $\bar{\omega}_{\mathrm{E}}$ for which the perturbation growth rate $\bar{e}^{-1}(\mathrm{d}\bar{e}/\mathrm{d}\bar{t})$ is a maximum. Now, the extremum of $\bar{e}^{-1}(\mathrm{d}\bar{e}/\mathrm{d}\bar{t})$ with respect to $\bar{\omega}$ follows from \eqref{eq:eDotOvere} as
\begin{align}\label{eq:eDotOvereMax}
    \bigg(\frac{1}{\bar{e}}\frac{\mathrm{d}\bar{e}}{\mathrm{d}\bar{t}}\bigg)_{\mathrm{E}} = 
    \frac{1}{\chi_1\chi_2^2}
    \bigg(\big[1+\chi_1\chi_2\big]^{\frac{1}{2}}-1\bigg)^2.
\end{align}
and occurs at $\bar{\omega}=\bar{\omega}_{\mathrm{E}}$, where
\begin{align}\label{eq:omgMax}
    \bar{\omega}_{\mathrm{E}}^2 
    = \chi_2^{-1}\bigg(\big[1+\chi_1\chi_2\big]^{\frac{1}{2}}-1\bigg).
\end{align}
Contours of $\bar{\omega}_{\mathrm{E}}$ are shown in $(\chi_1,\chi_2)$ space in figure \ref{fig:analyticalPerturbation}a. Note that the fastest growing wavenumber scales as $\bar{\omega}_{\mathrm{E}}=(\chi_1/2)^{1/2}$ for small $\chi_1$ or small $\chi_2$. The ratio $\bar{\omega}_{\mathrm{E}}/\bar{\omega}_{\mathrm{c}}$ follows directly from \eqref{eq:omgCrit} and \eqref{eq:omgMax} such that
\begin{align}
    \bigg(\frac{\bar{\omega}_{\mathrm{E}}}{\bar{\omega}_{\mathrm{c}}}\bigg)^2
    =
    \big(\chi_1\chi_2\big)^{-1}
    \bigg(\big[1+\chi_1\chi_2\big]^{\frac{1}{2}}-1\bigg),
\end{align}
as shown in figure \ref{fig:analyticalPerturbation}b. 

Conservation of mass between a spherical droplet of radius $R_{\mathrm{s}}$ and a wavelength $\lambda$ of wire requires $3\pi R_0^2\lambda = 4\pi R_{\mathrm{s}}^3$, and consequently the ratio $R_{\mathrm{s}}/R_0$ reads
\begin{align}\label{eq:rSphere}
    \bigg(\frac{R_{\mathrm{s}}}{R_0}\bigg)^6 =  \bigg(\frac{3\pi}{2}\bigg)^2 \chi_2\bigg(\big[1+\chi_1\chi_2\big]^{\frac{1}{2}}-1\bigg)^{-1},
\end{align}
upon making use of \eqref{eq:omgMax}. This dependence of $(R_{\mathrm{s}}/R_0)$ upon $(\chi_1,\chi_2)$ is illustrated in figure \ref{fig:analyticalPerturbation}c: $(R_{\mathrm{s}}/R_0)$ increases monotonically with increasing $\chi_2$ and decreasing $\chi_1$. Note that, for small $\chi_1$ or small $\chi_2$, \eqref{eq:rSphere} reduces to
\begin{align}
    \frac{R_{\mathrm{s}}}{R_0} =  
    \bigg(\frac{3\pi}{\sqrt{2}}\bigg)^{\frac{1}{3}} 
    \chi_1^{\frac{1}{6}},
\end{align}

Now consider the pinch-off time $\bar{t}_{\mathrm{p}}$. Full numerical simulations (discussed later) reveal that the pinch-off time is dominated by the initial stage of perturbation growth at small $\bar{e}$: consequently, the value of $(\bar{e}^{-1}(\mathrm{d}\bar{e}/\mathrm{d}\bar{t}))_{\mathrm{E}}$ can be used to estimate $\bar{t}_{\mathrm{p}}$. Integration of \eqref{eq:eDotOvereMax} from an initial imperfection of amplitude $\bar{e}_0$ to an arbitrary pinch-off value $\bar{e}_{\mathrm{p}}$ gives
\begin{align}\label{eq:tpPrediction}
    \bar{t}_{\mathrm{p}} =
    \chi_1\chi_2^2
    \bigg(\big[1+\chi_1\chi_2\big]^{\frac{1}{2}}-1\bigg)^{-2}
    \ln{\bigg(\frac{\bar{e}_{\mathrm{p}}}{\bar{e}_0}\bigg)}.
\end{align}
The logarithmic dependence of $\bar{t}_{\mathrm{p}}$ upon $\bar{e}_{\mathrm{p}}/\bar{e}_0$ implies that $\bar{t}_{\mathrm{p}}$ is relatively insensitive to the precise choice of $\bar{e}_0$ and $\bar{e}_{\mathrm{p}}$, but for definiteness we shall take $\bar{e}_0=10^{-3}$ and $\bar{e}_{\mathrm{p}}=0.8$ in the presentation of numerical results below. The formula \eqref{eq:tpPrediction} is shown in graphical form in figure \ref{fig:analyticalPerturbation}d; it simplifies to $\bar{t}_{\mathrm{p}}=4/\chi_1$ for small $(\chi_1,\chi_2)$.

\section{Full numerical study} 

The  above perturbation analysis considered the initial growth of a small imperfection. A full numerical solution is now obtained to study the shape evolution in both the initial and the later stages of shape evolution. The solution strategy builds upon that of \citet{Nichols1965}. 

\subsection{Numerical implementation}

We proceed to obtain a numerical solution to \eqref{eq:govEquVN}. First, we non-dimensionalise the problem as follows. Lengths are non-dimensionalised by the initial wire radius $R_0$, curvatures by $2R_0$, and the outward normal velocity $v_{\mathrm{n}}$ by $R_0^3/((1+c)\mathcal{D}\gamma_0)$ such that
\begin{align}
    \bar{x} = x/R_0; \quad \bar{y} = y/R_0; \quad \bar{S} = S/R_0; \quad \bar{\kappa}_i = 2R_0\kappa_i; \quad \bar{v}_{\mathrm{n}} = \frac{R_0^3 v_{\mathrm{n}}}{(1+c)\mathscr{D}\gamma_0}, 
\end{align}
where the subscript $i$ denotes $m,d,e$ for the mean, deviatoric, and effective curvature, respectively. Non-dimensionalise $\gamma$ and $\eta$ such that $\bar{\gamma}=(1+c)^{-1}(\gamma/\gamma_0)$ and $\bar{\eta}=\eta/\eta_0$, write $\kappa_0 = 1/(2R_0)$, and again limit attention to the case of a linear viscous interface reaction, i.e. $M=1$. Then, the governing equation \eqref{eq:govEquVN} reduces to
\begin{align}\label{eq:vnDimless}
    \bar{v}_{\mathrm{n}} 
    - \frac{1}{\bar{y}\Lambda}\frac{\partial}{\partial\bar{S}} \Bigg( \frac{\bar{y}}{\Lambda}\frac{\partial}{\partial\bar{S}} \Big[ \bar{\sigma}_{\mathrm{r0}} \bar{v}_{\mathrm{n}} 
    + \bar{\eta}\bar{\eta}_0 \bar{\kappa}_{\mathrm{m}}^2 \bar{v}_{\mathrm{n}} \Big] \Bigg)
    = \frac{1}{\bar{y}\Lambda}\frac{\partial}{\partial\bar{S}}\Bigg( \frac{\bar{y}}{\Lambda}\frac{\partial}{\partial\bar{S}}\Big[ \bar{\gamma}\bar{\kappa}_{\mathrm{m}} \Big] \Bigg).
\end{align}
Note that the dependent variables $\bar{y}$, $\bar{x}$, and $\theta$ can be described as a function of ($\bar{t}$, $\bar{S}$; $\bar{\sigma}_{\mathrm{r}0}$, $\bar{\eta}_0$, $\alpha$, $c$, $d$)
and the initial conditions ($\bar{Y}$, $\bar{X}$, $\Theta$). To simulate the evolution of wire geometry, we use a numerical scheme  based on finite differences in space and a forward Euler scheme in time, in similar manner to that of \citet{Nichols1965}. Our implementation differs from theirs as we employ a full Lagrangian formulation, as stated in equations \eqref{eq:stretchFactor} and \eqref{eq:govEquVN}. The scheme is summarised briefly in the following paragraph. 

The starting point of a simulation is the initial profile as parametrised by $\bar{y}(\bar{S})$ and $\theta(\bar{S})$, where the variable $\bar{S}$ is discretised into equidistant steps of value $\bar{h}$, such that $\bar{S}_i=i\bar{h}$ for $i=0,1,2,\ldots$. Periodic boundary conditions are enforced over the perturbation wavelength $\bar{\lambda}=2\pi/\bar{\omega}$. The simulation begins with the evaluation of the two principal curvatures, $\bar{\kappa}_1$ and $\bar{\kappa}_2$, by central differences. Central differences are also used to compute other spatial derivatives, as needed. For evaluation of equation \eqref{eq:vnDimless}, this gives a system of linear equations for $\bar{v}_{\mathrm{n}}$,  which are solved by a standard solver algorithm\footnote{Matlab function mldivide as implemented in Matlab 2017b}. The time derivative of the stretch ratio $\Lambda$ and of the inclination $\theta$ are given by 
\begin{align}
    \dot{\Lambda}(\bar{S}_i)
    = 
    -\bar{v}_{\mathrm{n}}\frac{\partial\theta}{\partial\bar{S}}
    \quad\text{and}\quad
    \dot{\theta}(\bar{S}_i) 
    = 
    \frac{1}{\Lambda}\frac{\partial\bar{v}_{\mathrm{n}}}{\partial\bar{S}},
\end{align}
respectively. The forward Euler method is used to update $\Lambda(\bar{S}_i)$ and $\theta(\bar{S}_i)$. The new profile is obtained by fitting a third-order Lagrange polynomial to the function $\Lambda\sin{\theta}$, and the integral
\begin{align}
    \bar{y}(\bar{S}_i) - \bar{y}(0) = \int_{0}^{\bar{S}_i} \Lambda\sin{(\theta)} \mathrm{d}\bar{S}',
\end{align}
is evaluated using a Newton-Cotes scheme. Periodic boundary conditions dictate that $\dot{\bar{y}}(0) = \bar{v}_{\mathrm{n}}(0)$. Equation \eqref{eq:vnDimless} is then solved in finite difference form, the geometry is updated over the time step, and the process is repeated. We note in passing that $\bar{x}(\bar{S}_i)$ is not required in the simulation but its value can be tracked in a similar manner to that of $\bar{y}(\bar{S}_i)$. Numerical stability is ensured by choosing a suitably small time increment. In agreement with the comments by \citet{Nichols1965}, instabilities appear for time increments of $\Delta\bar{t}\gtrsim0.8\times\bar{h}^4$. Throughout this study, the time increment is kept at $\Delta\bar{t}=0.4\times\bar{h}^4$.

\subsection{Prototypical results}

\begin{figure}[ht]
    \centering
    \includegraphics[width=1.0\textwidth]{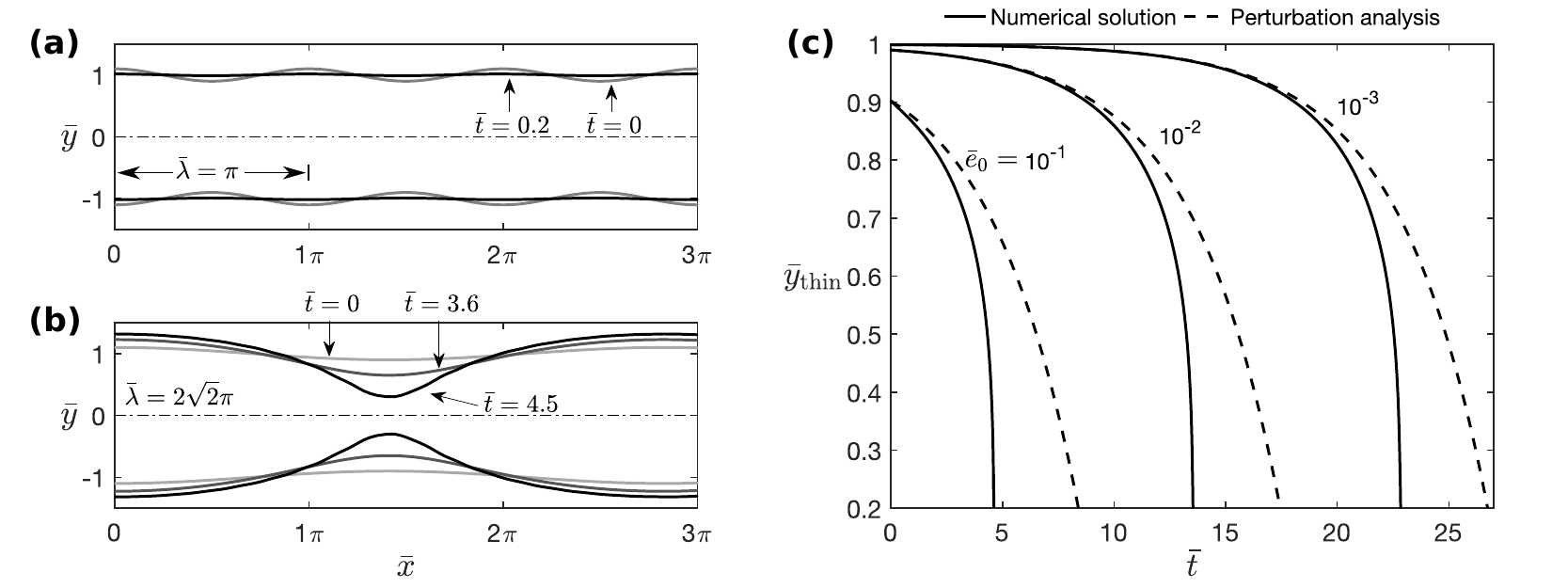}
    \caption{Shape evolution of a bare wire with $c=0$, $\bar{\sigma}_{\mathrm{r0}}=0$ and $\bar{\eta}_0=0$. (a) The perturbation decays for $\bar{\lambda} = \pi$, which is smaller than the critical wavelength $\bar{\lambda}_{\mathrm{c}} = 2\pi$. (b) The perturbation grows for $\bar{\lambda} = 2\sqrt{2}\pi > \bar{\lambda}_{\mathrm{c}}$. (c) Prediction of the time-evolution of the radius at the thinnest wire site, $\bar{y}_{\mathrm{min}}=\bar{y}(\bar{x}=\bar{\lambda}/2)$, for three choices of $\bar{e}_0$. Full numerical solution as solid line; perturbation analysis as dashed line.}
    \label{fig:shapeEvolutionBareWire}
\end{figure}

Checks were performed to ensure that the predictions of the perturbation analysis agree with the full numerical solution in the early stages of perturbation growth. For simplicity, we consider the reference case $c=\bar{\sigma}_{\mathrm{r0}}=\bar{\eta}_0=0$, corresponding to a nanowire absent the organic ligand shell, as analysed by \citet{Nichols1965}. Typical results for the shape evolution are given in figure \ref{fig:shapeEvolutionBareWire} for the choices $\bar{\lambda}=\pi$ and $\bar{\lambda}=2\sqrt{2}\pi$. Note that the critical perturbation wavelength $\bar{\lambda}_{\mathrm{c}}=2\pi/\bar{\omega}_{\mathrm{c}}$ takes a value of $2\pi$ in the present case. A stable response is obtained for $\bar{\lambda}=\pi<\bar{\lambda}_{\mathrm{c}}$, see figure \ref{fig:shapeEvolutionBareWire}a, whereas the perturbation grows for $\bar{\lambda}=2\sqrt{2}\pi>\bar{\lambda}_{\mathrm{c}}$, see figure \ref{fig:shapeEvolutionBareWire}b.

Now focus attention on the time-evolution of the thinnest section of the wire, $\bar{y}_{\mathrm{thin}}$, again for $c=\bar{\sigma}_{\mathrm{r0}}=\bar{\eta}_0=0$ and $\bar{\lambda}=2\sqrt{2}\pi$. The full numerical solution is compared with the perturbation analysis in figure \ref{fig:shapeEvolutionBareWire}c for selected values of initial perturbation amplitude $\bar{e}_0$. Recall that we have chosen to define the pinch-off time $\bar{t}_{\mathrm{p}}$ on the basis that $\bar{e}_{\mathrm{p}}=0.8$ such that $\bar{y}_{\mathrm{thin}}(\bar{t}_{\mathrm{p}})=1-\bar{e}_{\mathrm{p}}=0.2$. The predicted evolution of the thinnest section of the wire from the perturbation analysis is included in figure \ref{fig:shapeEvolutionBareWire}c. Excellent agreement is obtained in the initial stage of instability growth, but there is some divergence between perturbation theory and the full numerical prediction as the wire develops a deep notch at the pinch-off location. It is evident that $\bar{t}_{\mathrm{p}}$ scales as $-\ln{(\bar{e}_0)}$ for both the numerical analysis and the perturbation analysis, recall \eqref{eq:tpPrediction}. 

\subsection{Comparison of initial perturbation analysis and full numerical solution}

The perturbation analysis reveals that the early growth of the Rayleigh-Plateau instability is independent of the value of $d$ and is only controlled by $\chi_1$, which characterises the effective surface energy of wire and organic ligand shell, and by $\chi_2$, which characterises the kinetics of interface reaction and viscous drag. 

\begin{figure}[ht]
    \centering
    \includegraphics[width=1.0\textwidth]{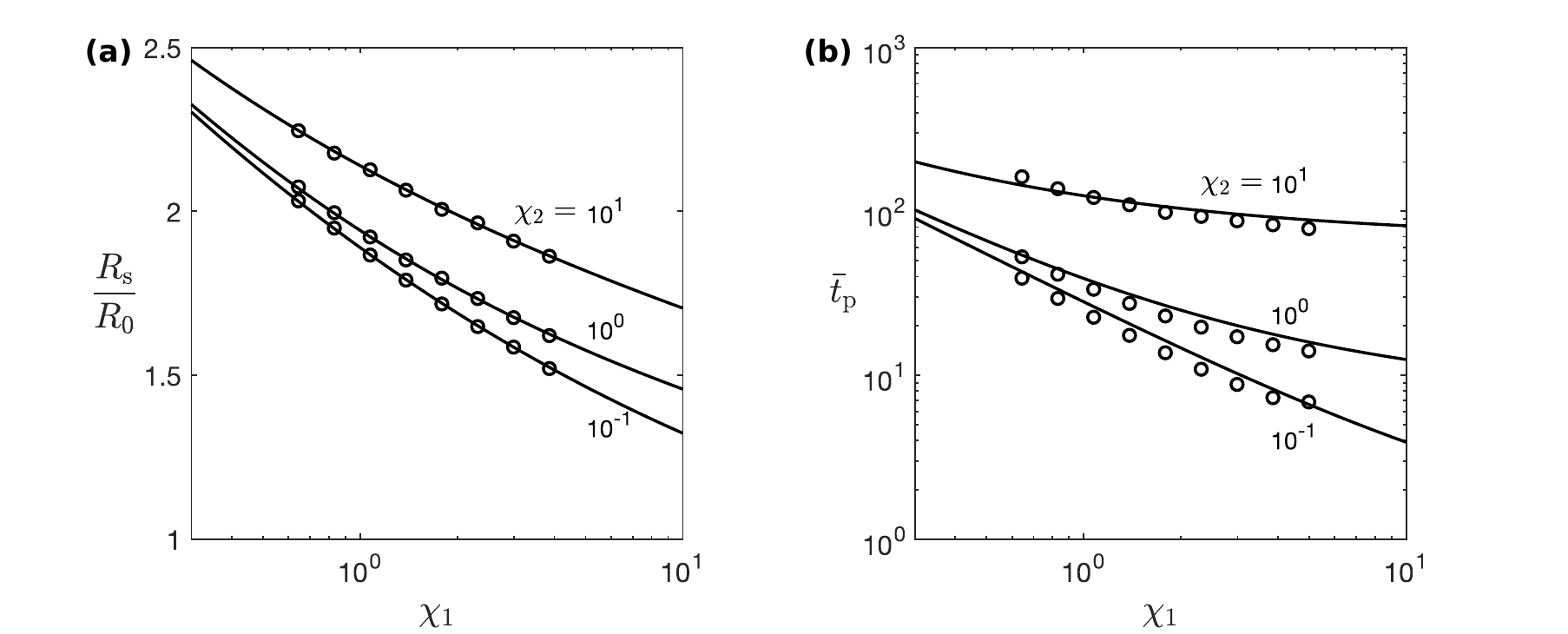}
    \caption{(a) Comparison of analytical solution \eqref{eq:rSphere} (solid line) and full numerical solution (data points) for the radius of spherical droplets;  (b) comparison of analytical solution \eqref{eq:tpPrediction} (solid line) and full numerical solution (data points) for the pinch-off time.}
    \label{fig:perturbationComparison}
\end{figure}

How accurate is the perturbation analysis in terms of prediction of the final droplet radius and of the pinch-off time? The predictions \eqref{eq:rSphere} for $R_{\mathrm{s}}/R_0$ and \eqref{eq:tpPrediction} for $\bar{t}_{\mathrm{p}}$ from the perturbation analysis are plotted in figure \ref{fig:perturbationComparison} using $\chi_1$ as the ordinate and for selected values of $\chi_2$. Full numerical results are included on the same plots for the choice $d=0$. Excellent agreement between the perturbation and full numerical analyses is noted for both $R_{\mathrm{s}}/R_0$ and $\bar{t}_{\mathrm{p}}$ implying that the main features of the instability are dictated by early growth of the imperfection, with $(\chi_1,\chi_2)$ playing the main role. Hence, the maps of figures \ref{fig:analyticalPerturbation}c and \ref{fig:analyticalPerturbation}d remain accurate beyond the initial growth phase of the instability. 

\subsection{Influence of the kinetic parameters upon morphology and pinch-off time}

\begin{figure}[ht]
    \centering
    \includegraphics[width=1.0\textwidth]{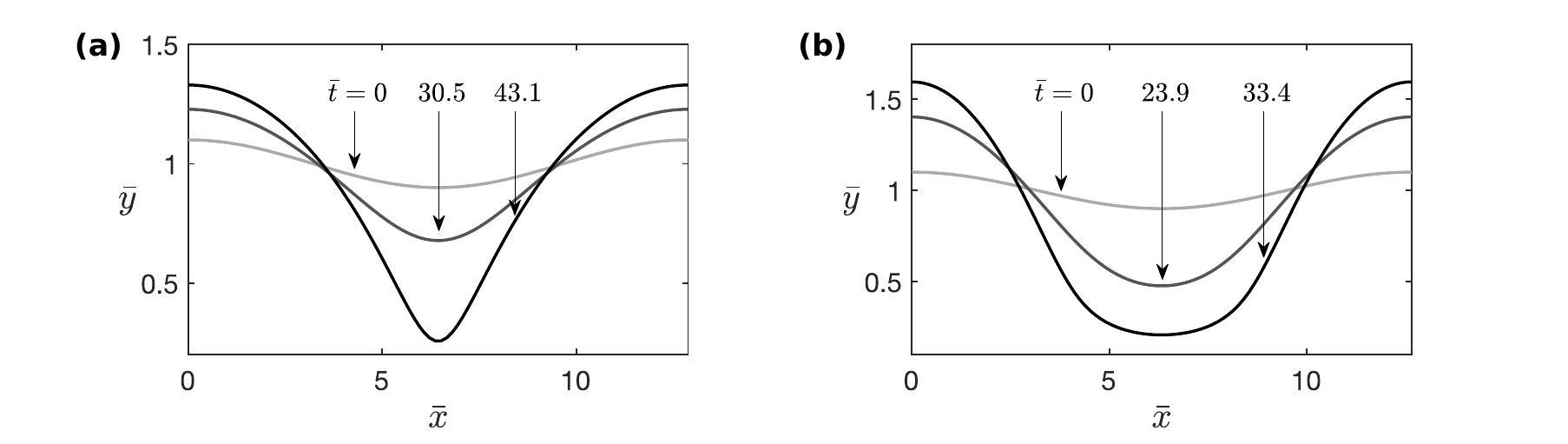}
    \caption{Sensitivity of the wire profile to the kinetic parameters $\bar{\sigma}_{\mathrm{r0}}$ and $\bar{\eta}_0$ for $(\chi_1,\chi_2)=(1,10)$ and  $(\alpha,c,d)=(0.5,0,0)$. (a) $\bar{\sigma}_{\mathrm{r0}}=10$ and $\bar{\eta}_0=0$. (b) $\bar{\sigma}_{\mathrm{r0}}=0$ and $\bar{\eta}_0=10$.}
    \label{fig:sensitivityKinetics}
\end{figure}

We anticipate that the later stages of the instability depend upon the various non-dimensional groups contained within $\chi_1$ and $\chi_2$. For example, consider $\chi_2=\bar{\sigma}_{\mathrm{r0}}+\bar{\eta}_0$. What if we keep $\chi_2$ fixed and vary the ratio $\bar{\sigma}_{\mathrm{r0}}/\bar{\eta}_0$? To explore this, we have performed a full numerical simulation for the case $(\bar{\sigma}_{\mathrm{r0}},\bar{\eta}_0)$ equal to $(10,0)$ and then equal to $(0,10)$ such that $\chi_2=10$ in both simulations. The other parameters were held fixed at $\alpha=0.5$, $c=0$, and $d=0$. As anticipated, the early growth of the instability is the same in both cases, see figure \ref{fig:sensitivityKinetics}. But at later times, the pinch-off shape is sharper for the choice $\bar{\sigma}_{\mathrm{r0}}=10$ and $\bar{\eta}_0=0$ (figure \ref{fig:sensitivityKinetics}a) than for the other choice (figure \ref{fig:sensitivityKinetics}b). 

\begin{figure}[ht]
    \centering
    \includegraphics[width=1.0\textwidth]{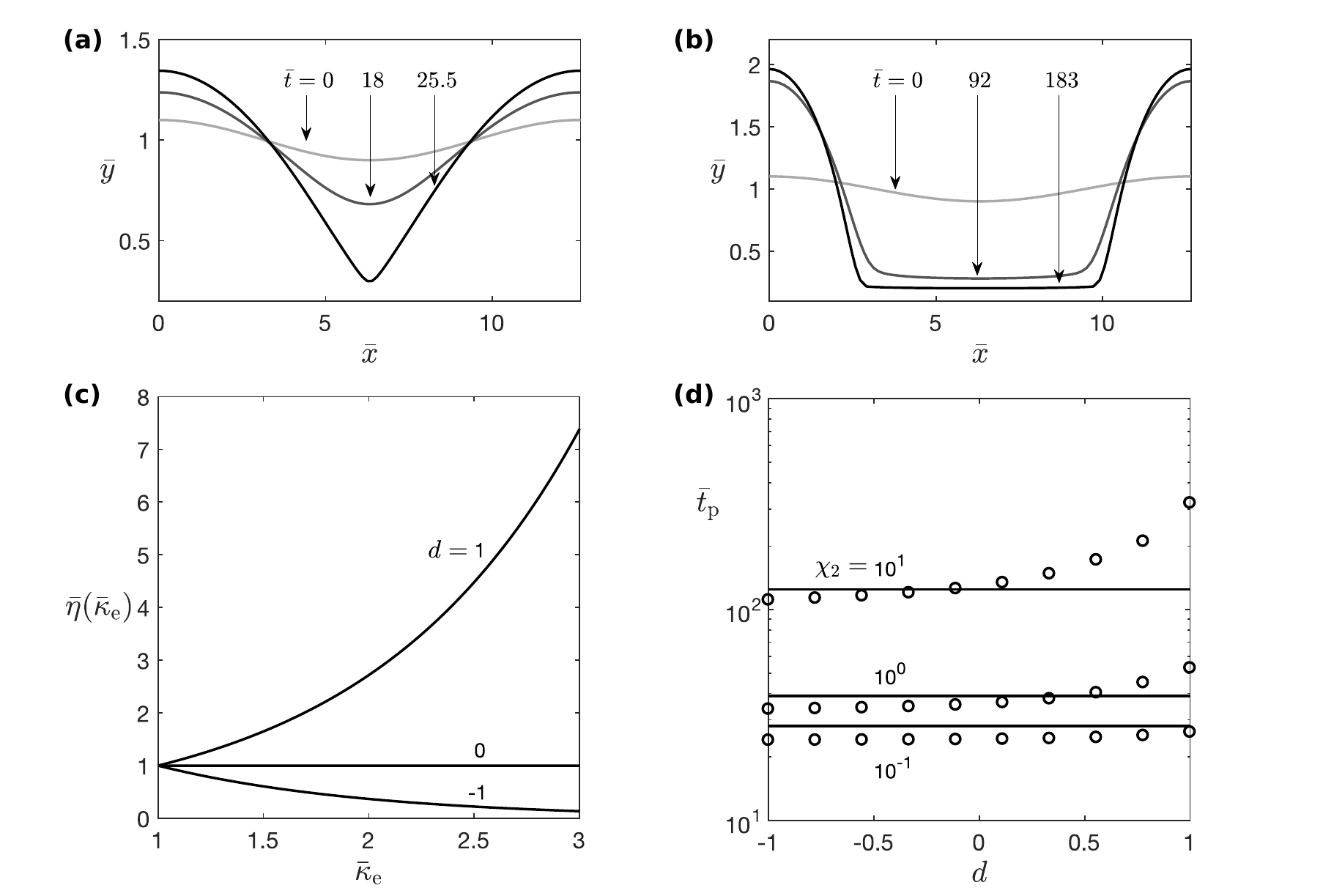}
    \caption{(a) Wire evolution for $(\chi_1,\chi_2)=(1,10)$ and $d=-1$. (b) Wire evolution for $(\chi_1,\chi_2)=(1,10)$ and $d=+1$. (c) Sensitivity of $\bar{\eta}_0$ to $d$ for increasing effective curvature $\bar{\kappa}_{\mathrm{e}}$. (d) Sensitivity of the pinch-off time $\bar{t}_{\mathrm{p}}$ to $d$ for $\chi_1=1$ and $\chi_2=(0.1, 1, 10)$. The prediction by the initial perturbation analysis is shown as a solid line and is insensitive to $d$. The numerical results are shown as circles.}
    \label{fig:dSensitivity}
\end{figure}

It remains to explore the role of $d$ in influencing the instability. Consider the evolution of wire shape for the choice $d=-1$ in figure \ref{fig:dSensitivity}a and for $d=1$ in figure \ref{fig:dSensitivity}b. In both cases, we take $\alpha=0.5$, $c=0$, $\bar{\sigma}_{\mathrm{r0}}=0$, and $\bar{\eta}_0=10$ such that $\chi_1=1$ and $\chi_2=10$. (We note in passing that the plot in figure \ref{fig:sensitivityKinetics}b is for the same parameter values, but with $d=0$.) A sharp notch develops in the profile for $d=-1$, whereas for $d=1$ the degree of viscous drag within the organic ligand shell has a strong stabilising influence, and the wire adopts a uniform high curvature $\kappa_1$ (with $\kappa_2\approx 0$) over a significant portion of the wire. To gain further insight, we have plotted $\bar{\eta}$ versus $\bar{\kappa}_{\mathrm{e}}$, as defined in equation \eqref{eq:etaFunctional}, in figure \ref{fig:dSensitivity}c. The choice of $d=1$ leads to a steep increase in $\bar{\eta}$ with increasing curvature $\bar{\kappa}_{\mathrm{e}}$, thereby stabilising the wire against continued pinch-off. The dependence of $\bar{t}_{\mathrm{p}}$ upon $d$ is shown explicitly in figure \ref{fig:dSensitivity}d for $\chi_1=1$ and for three selected values of $\chi_2$. Consider the case $\chi_2=10$, as discussed in reference to figures \ref{fig:dSensitivity}a and \ref{fig:dSensitivity}b. As $d$ increases from -1 to 1, there is a moderate increase in $\bar{t}_{\mathrm{p}}$ due to the increase in $\bar{\eta}$ at high local curvature. 

\section{Concluding discussion}

The present study reveals the significant role played by an organic ligand shell in the Rayleigh-Plateau instability for nanowires. It assumes an axisymmetric response which is appropriate for isotropic behaviour. Additionally, functional forms for the surface energy and the viscosity are assumed and a measure for the effective curvature is introduced. We anticipate that the general behaviour of the organic ligand shell can be adequately captured by the effective curvature and these functionals. Future modelling at the molecular length scale can give further insight into appropriate continuum descriptions. 

\begin{figure}[ht]
    \centering
    \includegraphics[width=1.0\textwidth]{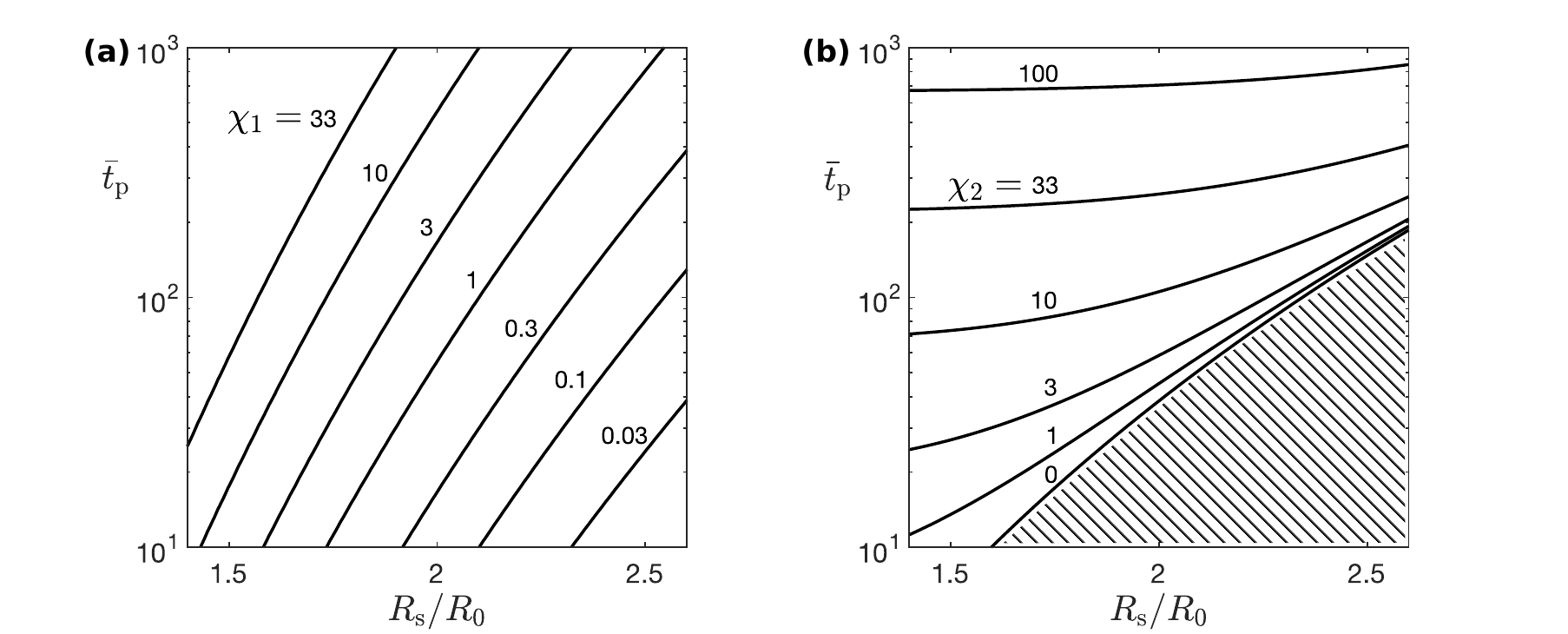}
    \caption{The inverse problem of extraction of the magnitude of (a) $\chi_1$ and (b) $\chi_2$ from given values of $R_{\mathrm{s}}/R_0$ and $\bar{t}_{\mathrm{p}}$.}
    \label{fig:chi1chi2Prediction}
\end{figure}

Before summarising how our findings can guide the development of geometrically stable nanowires, we shall first show how our findings can be used to broaden our understanding of the Rayleigh-Plateau instability of currently available nanowires. The maps of figure \ref{fig:analyticalPerturbation} highlight the dependence of pinch-off time $\bar{t}_{\mathrm{p}}$ and final droplet radius $R_{\mathrm{s}}/R_0$ upon the two dominant non-dimensional groups $\chi_1$ and $\chi_2$. In order to deduce information about currently available nanowires, it is instructive to replot these maps as contours of $\chi_1$ and $\chi_2$ with $R_{\mathrm{s}}/R_0$ and $\bar{t}_{\mathrm{p}}$ as axes, see figure \ref{fig:chi1chi2Prediction}. We note in passing the formulae \eqref{eq:rSphere} and \eqref{eq:tpPrediction} can be inverted algebraically to give
\begin{align}
    \chi_1 
    = 
    \bigg( \frac{3\pi}{2}\bigg)^4 
    \bigg(\ln{\bigg(\frac{\bar{e}_{\mathrm{p}}}{\bar{e}_0}\bigg)}\bigg)^{-1}
    \bigg(\frac{R_{\mathrm{s}}}{R_0}\bigg)^{-12}
    \bar{t}_{\mathrm{p}}
\end{align}
and 
\begin{align}
    \chi_2
    = 
    \bigg(\ln{\bigg(\frac{\bar{e}_{\mathrm{p}}}{\bar{e}_0}\bigg)}\bigg)^{-1}
    \bar{t}_{\mathrm{p}}
    - \bigg(\frac{2\sqrt{2}}{3\pi}\bigg)^2
    \bigg(\frac{R_{\mathrm{s}}}{R_0}\bigg)^{6},
\end{align}
as depicted in figure \ref{fig:chi1chi2Prediction}. Recall that both $\chi_1$ and $\chi_2$ are non-negative. This restricts the contour plot of $\chi_2$ in figure \ref{fig:chi1chi2Prediction}b as illustrated by the shaded region. The revised map of figure \ref{fig:chi1chi2Prediction}a may be used to deduce information about $\chi_1$ and thereby the energetics of the organic ligand shell from measurements of $R_{\mathrm{s}}/R_0$ and $\bar{t}_{\mathrm{p}}$. Likewise, the map of figure \ref{fig:chi1chi2Prediction}b can be used to deduce a value for $\chi_2$, and consequently the kinetics of the organic ligand shell, again from observed values of $R_{\mathrm{s}}/R_0$ and $\bar{t}_{\mathrm{p}}$. 

Our findings can guide the selection of organic ligands that render nanowires more stable. The above results confirm that nanowires can be stabilised energetically against break-up into discrete droplets, and the relevant parameter range in $(\alpha,c)$ space is given in figure  \ref{fig:analyticalRegimeMap}. Stabilisation requires the combined surface energy of wire and organic ligand shell to decrease with increasing effective curvature. When break-up is inevitable on energetic grounds, there is still the opportunity to delay its occurrence by judicious control of the kinetics of diffusion, of the interface reaction, and of the viscous dissipation within the organic ligand shell. Suitable ligands should thus minimise surface energy, maximise the viscous dissipation captured by the non-dimensional group $\chi_2$, and cause an increase in the shell's viscosity with increasing effective curvature. Molecular dynamics simulations and systematic experimentation can help to find suitable molecules.

\section{Acknowledgements}

The authors are grateful for financial support of this work in the form of an ERC \mbox{MULTILAT} grant 669764, and to the Humboldt Society (NAF was the recipient of a Humboldt award). SB gratefully acknowledges the Cambridge Centre for Micromechanics, the INM, and the German Academic Exchange Service for funding a number of research stays at the University of Cambridge. SB and TK would like to thank Eduard Arzt for his continuing support of the project. The authors would like to thank Louis V. Weber and Lola Gonz\'alez-Garc\'ia for recording TEM micrographs of the nanowires. 

\newpage

\section{Appendix: Determination of viscous drag stress}

Consider a metallic sphere of radius $R$ and a surrounding organic ligand shell of thickness $H$. A change in radius at rate $\dot{R}$ leads to a hoop strain ate $\dot{\varepsilon}_{\mathrm{hoop}}$ of the shell and, in turn, to viscous dissipation within the shell due to its shear viscosity $\eta$. Upon writing $v_{\mathrm{n}}=\dot{R}$ the hoop strain rate is
\begin{equation}\label{eq:hoopStrainRate}
    \dot\varepsilon_{\mathrm{hoop}} = \frac{v_{\mathrm{n}}}{R}
\end{equation}
The von Mises strain rate $\dot{\varepsilon}_{\mathrm{e}}$ reads
\begin{align}
    \dot\varepsilon_{\mathrm{e}} = \frac{2}{\sqrt{3}}\lvert\dot{\varepsilon}_{\mathrm{hoop}}\rvert
\end{align}
and the von Mises stress $\sigma_{\mathrm{e}}$ is related to $\dot{\varepsilon}_{\mathrm{e}}$ by the normal relation 
\begin{equation}
    \sigma_{\mathrm{e}} = 3\eta\dot\varepsilon_{\mathrm{e}},
\end{equation}
in terms of a shear viscosity $\eta$. The work rate per unit volume is given by
\begin{equation}\label{eq:workRateCreep}
    \sigma_{\mathrm{e}}\dot\varepsilon_{\mathrm{e}} = 3\eta\dot\varepsilon_{\mathrm{e}}^2 = 4\eta\dot\varepsilon_{\mathrm{hoop}}^2,
\end{equation}
and the dissipation potential per unit area is
\begin{equation}\label{eq:dissPotential}
    \phi_{\eta} = \frac{1}{2} H \sigma_{\mathrm{e}}\dot\varepsilon_{\mathrm{e}} = 2 H \eta \Big( \frac{v_{\mathrm{n}}}{R} \Big)^2.
\end{equation}
Differentiate \eqref{eq:dissPotential} with respect to $v_\mathrm{n}$ to obtain the viscous drag stress
\begin{align}
    \sigma_{\eta} = \frac{\partial \phi_{\eta}}{\partial v_{\mathrm{n}}} = 4 H \eta \kappa_{\mathrm{m}}^2 v_{\mathrm{n}}
\end{align}
upon recalling that $\kappa_{\mathrm{m}}=1/R$ for a sphere.

\newpage
\section{List of symbols}

\begin{table}[!htbp]
    \scriptsize
    \begin{tabular}{c l}
        Variable & Description (units) \\ \hline
        $\alpha$ & \enquote{Shape} parameter in effective curvature (-) \\
        $\gamma$ & Surface energy (N/m) \\
        $\gamma_0$ & Material parameter in surface energy (N/m) \\ 
        $\delta_{\mathrm{b}}$ & Thickness of boundary layer (m) \\
        $\dot{\varepsilon}$ & Strain rate ($\mathrm{s^{-1}}$) \\
        $\dot{\varepsilon}_{\mathrm{e}}$ & Von Mises strain rate ($\mathrm{s^{-1}}$) \\
        $\dot{\varepsilon}_{\mathrm{hoop}}$ & Hoop strain rate ($\mathrm{s^{-1}}$) \\
        $\eta$ & Viscosity ($\mathrm{Nm^{-2}s}$) \\
        $\eta_0$ & Material parameter in viscosity ($\mathrm{Nm^{-2}s}$) \\
        $\theta$ & Inclination of surface in current configuration (-) \\
        $\Theta$ & Inclination of surface in initial configuration (-) \\
        $\kappa_1$ & Principal curvature in $1$-direction ($\mathrm{m^{-1}}$) \\
        $\kappa_2$ & Principle curvature in $2$-direction ($\mathrm{m^{-1}}$) \\
        $\kappa_{\mathrm{m}}$ & Mean curvature ($\mathrm{m^{-1}}$) \\
        $\kappa_{\mathrm{d}}$ & Deviatoric curvature ($\mathrm{m^{-1}}$) \\
        $\kappa_{\mathrm{e}}$ & Effective curvature ($\mathrm{m^{-1}}$) \\
        $\lambda$ & Perturbation wavelength ($\mathrm{m}$) \\
        $\lambda_{\mathrm{c}}$ & Critical perturbation wavelength ($\mathrm{m}$) \\
        $\Lambda$ & Stretch ratio of surface (-) \\
        $\mu$ & Chemical potential per atom ($\mathrm{J}$) \\
        $\sigma_{\mathrm{r}}$ & Interface reaction stress ($\mathrm{Nm^{-2}}$) \\
        $\sigma_{\mathrm{r0}}$ & Material parameter in interface reaction stress ($\mathrm{Nm^{-2}}$) \\
        $\sigma_{\eta}$ & Viscous drag stress ($\mathrm{Nm^{-2}}$) \\
        $\chi_1$ & Non-dimensional group containing energetic terms (-) \\
        $\chi_2$ & Non-dimensional group containing dissipative terms (-) \\
        $\omega$ & Perturbation wavenumber ($\mathrm{m^{-1}}$) \\
        $\omega_{\mathrm{c}}$ & Critical perturbation wavenumber ($\mathrm{m^{-1}}$) \\
        $\omega_{\mathrm{E}}$ & Perturbation wavenumber of fastest growing perturbation ($\mathrm{m^{-1}}$) \\
        $\Omega$ & Atomic volume ($\mathrm{m^3}$) \\
        $c$ & Material constant in surface energy (-) (equation \ref{eq:gammaFunctional})\\
        $d$ & Material constant in viscosity (-) (equation \ref{eq:etaFunctional})\\
        $D_{\mathrm{b}}$ & Interface diffusivity ($\mathrm{m^2s^{-1}}$) \\
        $\mathscr{D}$ & Interface diffusion constant ($\mathrm{J^{-1}m^6s^{-1}}$) \\
        $e$ & Perturbation amplitude ($\mathrm{m}$) \\
        $e_0$ & Initial perturbation amplitude ($\mathrm{m}$) \\
        $e_{\mathrm{p}}$ & Perturbation amplitude at pinch-off ($\mathrm{m}$) \\
        $f$ & Driving force for kinetic dissipation ($\mathrm{N}$) \\
        $f_{\mathrm{r}}$ & Driving force for interface reaction ($\mathrm{N}$) \\
        $f_{\mathrm{\eta}}$ & Driving force for viscous dissipation ($\mathrm{N}$) \\
        $h$ & Step size in finite difference scheme ($\mathrm{m}$) \\
        $H$ & Layer thickness of the organic ligand shell ($\mathrm{m}$) \\
        $j$ & Surface diffusion flux ($\mathrm{m^2s^{-1}}$) \\
        $k$ & Boltzmann constant ($\mathrm{JK^{-1}}$) \\
        $M$ & Power law exponent in interface reaction (-) \\
        $R$ & Wire radius ($\mathrm{m}$) \\
        $R_0$ & Initial wire radius ($\mathrm{m}$) \\
        $R_{\mathrm{s}}$ & Radius of spherical droplets ($\mathrm{m}$) \\
        $s$ & Arc length coordinate ($\mathrm{m}$) \\
        $S$ & Arc length coordinate in the initial configuration ($\mathrm{m}$) \\
        $t$ & Time ($\mathrm{s}$) \\
        $t_{\mathrm{p}}$ & Pinch-off time ($\mathrm{s}$) \\
        $T$ & Absolute temperature ($\mathrm{K}$) \\
        $v_{\mathrm{n}}$ & Outward normal velocity of surface ($\mathrm{ms^{-1}}$) \\
        $v_{\mathrm{r}0}$ & Reference velocity in interface reaction ($\mathrm{ms^{-1}}$) \\
        $x$ & Cylindrical coordinate along wire-axis ($\mathrm{m}$) \\
        $X$ & Cylindrical coordinate along wire-axis in initial configuration ($\mathrm{m}$) \\
        $y$ & Cylindrical coordinate perpendicular to wire-axis ($\mathrm{m}$) \\
        $Y$ & Cylindrical coordinate perpendicular to wire-axis in initial configuration ($\mathrm{m}$) \\
        - & Non-dimensional versions of symbols are marked with a bar on top of the original symbol. \\
    \end{tabular}
    \label{tab:symbols}
\end{table}



\bibliographystyle{elsarticle-harv}

\newpage
\bibliography{references}

\begin{thebibliography}{47}
\expandafter\ifx\csname natexlab\endcsname\relax\def\natexlab#1{#1}\fi
\expandafter\ifx\csname url\endcsname\relax
  \def\url#1{\texttt{#1}}\fi
\expandafter\ifx\csname urlprefix\endcsname\relax\def\urlprefix{URL }\fi

\bibitem[{Ashby(1969)}]{Ashby1969}
Ashby, M., nov 1969. {On interface-reaction control of Nabarro-Herring creep
  and sintering}. Scripta Metallurgica 3~(11), 837--842.
\newline\urlprefix\url{http://linkinghub.elsevier.com/retrieve/pii/0036974869901914}

\bibitem[{Barwicz et~al.(2012)Barwicz, Cohen, Reuter, Bangsaruntip, and
  Sleight}]{Barwicz2012}
Barwicz, T., Cohen, G.~M., Reuter, K.~B., Bangsaruntip, S., Sleight, J.~W., feb
  2012. {Anisotropic capillary instability of silicon nanostructures under
  hydrogen anneal}. Applied Physics Letters 100~(9), 093109.
\newline\urlprefix\url{http://aip.scitation.org/doi/10.1063/1.3690869}

\bibitem[{Bid et~al.(2005)Bid, Bora, and Raychaudhuri}]{Bid2005}
Bid, A., Bora, A., Raychaudhuri, A.~K., may 2005. {Experimental study of
  Rayleigh instability in metallic nanowires using resistance fluctuations
  measurements from 77K to 375K}. In: Svedlindh, P., Popovic, D., Weissman,
  M.~B. (Eds.), Fluctuations and Noise in Materials II. Vol. 5843. p. 147.
\newline\urlprefix\url{http://proceedings.spiedigitallibrary.org/proceeding.aspx?doi=10.1117/12.609419}

\bibitem[{Cademartiri et~al.(2008)Cademartiri, Malakooti, O'Brien, Migliori,
  Petrov, Kherani, and Ozin}]{Cademartiri2008}
Cademartiri, L., Malakooti, R., O'Brien, P.~G., Migliori, A., Petrov, S.,
  Kherani, N.~P., Ozin, G.~A., may 2008. {Large-Scale Synthesis of Ultrathin
  Bi2S3 Necklace Nanowires}. Angewandte Chemie International Edition 47~(20),
  3814--3817.
\newline\urlprefix\url{http://doi.wiley.com/10.1002/anie.200705034}

\bibitem[{Cademartiri and Ozin(2009)}]{Cademartiri2009}
Cademartiri, L., Ozin, G.~A., mar 2009. {Ultrathin Nanowires-A Materials
  Chemistry Perspective}. Advanced Materials 21~(9), 1013--1020.
\newline\urlprefix\url{http://doi.wiley.com/10.1002/adma.200801836}

\bibitem[{Chen et~al.(2013)Chen, Ouyang, Gu, and Cheng}]{Chen2013}
Chen, Y., Ouyang, Z., Gu, M., Cheng, W., jan 2013. {Mechanically Strong,
  Optically Transparent, Giant Metal Superlattice Nanomembranes From Ultrathin
  Gold Nanowires}. Advanced Materials 25~(1), 80--85.
\newline\urlprefix\url{http://doi.wiley.com/10.1002/adma.201202241}

\bibitem[{Ciuculescu et~al.(2009)Ciuculescu, Dumestre, Comesana-Hermo,
  Chaudret, Spasova, Farle, and Amiens}]{Ciuculescu2009}
Ciuculescu, D., Dumestre, F., Comesana-Hermo, M., Chaudret, B., Spasova, M.,
  Farle, M., Amiens, C., sep 2009. {Single-Crystalline Co Nanowires: Synthesis,
  Thermal Stability, and Carbon Coating}. Chemistry of Materials 21~(17),
  3987--3995.
\newline\urlprefix\url{http://pubs.acs.org/doi/abs/10.1021/cm901349y}

\bibitem[{Cocks(1992)}]{Cocks1992}
Cocks, A., apr 1992. {Interface reaction controlled creep}. Mechanics of
  Materials 13~(2), 165--174.
\newline\urlprefix\url{http://linkinghub.elsevier.com/retrieve/pii/016766369290044E}

\bibitem[{Cocks et~al.(1998)Cocks, Gill, and Pan}]{Cocks1998}
Cocks, A. C.~F., Gill, S. P.~A., Pan, J., 1998. {Modeling Microstructure
  Evolution in Engineering Materials}. In: van~der Giessen, E., Wu, T. (Eds.),
  Advances in Applied Mechanics Volume 36. Academic Press, pp. 82--162.
\newline\urlprefix\url{https://www.elsevier.com/books/advances-in-applied-mechanics/unknown/978-0-12-002036-2}

\bibitem[{Feng et~al.(2009)Feng, Yang, You, Li, Guo, Yu, Shen, Wu, and
  Xing}]{Feng2009}
Feng, H., Yang, Y., You, Y., Li, G., Guo, J., Yu, T., Shen, Z., Wu, T., Xing,
  B., 2009. {Simple and rapid synthesis of ultrathin gold nanowires, their
  self-assembly and application in surface-enhanced Raman scattering}. Chemical
  Communications 0~(15), 1984.
\newline\urlprefix\url{http://xlink.rsc.org/?DOI=b822507a}

\bibitem[{Frost and Ashby(1982)}]{Frost1982}
Frost, H.~J., Ashby, M., 1982. {Deformation-mechanism maps: the plasticity and
  creep of metals and ceramics}. Pergamon Press, Oxford, New York, Sydney.

\bibitem[{Gong et~al.(2014)Gong, Schwalb, Wang, Chen, Tang, Si, Shirinzadeh,
  and Cheng}]{Gong2014}
Gong, S., Schwalb, W., Wang, Y., Chen, Y., Tang, Y., Si, J., Shirinzadeh, B.,
  Cheng, W., dec 2014. {A wearable and highly sensitive pressure sensor with
  ultrathin gold nanowires}. Nature Communications 5~(1), 3132.
\newline\urlprefix\url{http://www.nature.com/articles/ncomms4132}

\bibitem[{Hopwood and Mann(1997)}]{Hopwood1997}
Hopwood, J.~D., Mann, S., aug 1997. {Synthesis of Barium Sulfate Nanoparticles
  and Nanofilaments in Reverse Micelles and Microemulsions}. Chemistry of
  Materials 9~(8), 1819--1828.
\newline\urlprefix\url{http://pubs.acs.org/doi/abs/10.1021/cm970113q}

\bibitem[{Huang et~al.(2010)Huang, Zhan, Wang, Zhang, Xing, Guo, Leusink,
  Zheng, and Wu}]{Huang2010}
Huang, X.~H., Zhan, Z.~Y., Wang, X., Zhang, Z., Xing, G.~Z., Guo, D.~L.,
  Leusink, D.~P., Zheng, L.~X., Wu, T., nov 2010. {Rayleigh-instability-driven
  simultaneous morphological and compositional transformation from Co nanowires
  to CoO octahedra}. Applied Physics Letters 97~(20), 203112.
\newline\urlprefix\url{http://aip.scitation.org/doi/10.1063/1.3518470}

\bibitem[{Huber et~al.(2012)Huber, Warakulwit, Limtrakul, Tsukuda, and
  Probst}]{Huber2012}
Huber, S.~E., Warakulwit, C., Limtrakul, J., Tsukuda, T., Probst, M., 2012.
  {Thermal stabilization of thin gold nanowires by surfactant-coating: a
  molecular dynamics study}. Nanoscale 4~(2), 585--590.
\newline\urlprefix\url{http://xlink.rsc.org/?DOI=C1NR11282A}

\bibitem[{ITRS(2011)}]{ITRS2011}
ITRS, 2011. {International Technology Roadmap for Semiconductors. ITRS 2011
  Edition.} Tech. rep., World Semiconductor Council.
\newline\urlprefix\url{http://www.itrs2.net/2011-itrs.html}

\bibitem[{Karim et~al.(2006)Karim, Toimil-Molares, Balogh, Ensinger, Cornelius,
  Khan, and Neumann}]{Karim2006}
Karim, S., Toimil-Molares, M.~E., Balogh, A.~G., Ensinger, W., Cornelius,
  T.~W., Khan, E.~U., Neumann, R., dec 2006. {Morphological evolution of Au
  nanowires controlled by Rayleigh instability}. Nanotechnology 17~(24),
  5954--5959.
\newline\urlprefix\url{http://stacks.iop.org/0957-4484/17/i=24/a=009?key=crossref.ace7f89b491888e67281fa4095ad9a42}

\bibitem[{Lacroix et~al.(2014)Lacroix, Arenal, and Viau}]{Lacroix2014}
Lacroix, L.-M., Arenal, R., Viau, G., sep 2014. {Dynamic HAADF-STEM Observation
  of a Single-Atom Chain as the Transient State of Gold Ultrathin Nanowire
  Breakdown}. Journal of the American Chemical Society 136~(38), 13075--13077.
\newline\urlprefix\url{http://pubs.acs.org/doi/10.1021/ja507728j}

\bibitem[{Li et~al.(2015)Li, Ye, Stewart, Alvarez, and Wiley}]{Li2015}
Li, B., Ye, S., Stewart, I.~E., Alvarez, S., Wiley, B.~J., oct 2015. {Synthesis
  and Purification of Silver Nanowires To Make Conducting Films with a
  Transmittance of 99{\%}}. Nano Letters 15~(10), 6722--6726.
\newline\urlprefix\url{http://pubs.acs.org/doi/10.1021/acs.nanolett.5b02582}

\bibitem[{Liu et~al.(2005)Liu, Xu, Liang, Shen, Zhang, and Qian}]{Liu2005}
Liu, Z., Xu, D., Liang, J., Shen, J., Zhang, S., Qian, Y., jun 2005. {Growth of
  Cu 2 S Ultrathin Nanowires in a Binary Surfactant Solvent}. The Journal of
  Physical Chemistry B 109~(21), 10699--10704.
\newline\urlprefix\url{http://pubs.acs.org/doi/abs/10.1021/jp050332w}

\bibitem[{Lu et~al.(2008)Lu, Yavuz, Tuan, Korgel, and Xia}]{Lu2008}
Lu, X., Yavuz, M.~S., Tuan, H.-y., Korgel, B.~A., Xia, Y., jul 2008. {Ultrathin
  Gold Nanowires Can Be Obtained by Reducing Polymeric Strands of
  Oleylamine−AuCl Complexes Formed via Aurophilic Interaction}. Journal of
  the American Chemical Society 130~(28), 8900--8901.
\newline\urlprefix\url{http://pubs.acs.org/doi/abs/10.1021/ja803343m}

\bibitem[{Malakooti et~al.(2008)Malakooti, Cademartiri, Migliori, and
  Ozin}]{Malakooti2008}
Malakooti, R., Cademartiri, L., Migliori, A., Ozin, G.~A., 2008. {Ultrathin Sb
  2 S 3 nanowires and nanoplatelets}. J. Mater. Chem. 18~(1), 66--69.
\newline\urlprefix\url{http://xlink.rsc.org/?DOI=B713383A}

\bibitem[{Maurer et~al.(2015)Maurer, Gonz{\'{a}}lez-Garc{\'{i}}a, Reiser,
  Kanelidis, and Kraus}]{Maurer2015}
Maurer, J. H.~M., Gonz{\'{a}}lez-Garc{\'{i}}a, L., Reiser, B., Kanelidis, I.,
  Kraus, T., apr 2015. {Sintering of Ultrathin Gold Nanowires for Transparent
  Electronics}. ACS Applied Materials {\&} Interfaces 7~(15), 7838--7842.
\newline\urlprefix\url{http://pubs.acs.org/doi/10.1021/acsami.5b02088}

\bibitem[{Maurer et~al.(2016)Maurer, Gonz{\'{a}}lez-Garc{\'{i}}a, Reiser,
  Kanelidis, and Kraus}]{Maurer2016}
Maurer, J. H.~M., Gonz{\'{a}}lez-Garc{\'{i}}a, L., Reiser, B., Kanelidis, I.,
  Kraus, T., may 2016. {Templated Self-Assembly of Ultrathin Gold Nanowires by
  Nanoimprinting for Transparent Flexible Electronics}. Nano Letters 16~(5),
  2921--2925.
\newline\urlprefix\url{http://pubs.acs.org/doi/10.1021/acs.nanolett.5b04319}

\bibitem[{Nichols and Mullins(1965{\natexlab{a}})}]{Nichols1965}
Nichols, F.~A., Mullins, W.~W., jun 1965{\natexlab{a}}. {Morphological Changes
  of a Surface of Revolution due to Capillarity‐Induced Surface Diffusion}.
  Journal of Applied Physics 36~(6), 1826--1835.
\newline\urlprefix\url{http://aip.scitation.org/doi/10.1063/1.1714360}

\bibitem[{Nichols and Mullins(1965{\natexlab{b}})}]{Nichols1965AIME}
Nichols, F.~A., Mullins, W.~W., oct 1965{\natexlab{b}}. {Surface- (interface-)
  and volume diffusion contributions to morphological changes driven by
  capillarity}. AIME Metallurgical Society Transactions 233~(10), 1840--1848.
\newline\urlprefix\url{http://www.onemine.org/document/abstract.cfm?docid=26784}

\bibitem[{Pazos-Perez et~al.(2008)Pazos-Perez, Baranov, Irsen, Hilgendorff,
  Liz-Marzan, and Giersig}]{PazosPerez2008}
Pazos-Perez, N., Baranov, D., Irsen, S., Hilgendorff, M., Liz-Marzan, L.~M.,
  Giersig, M., sep 2008. {Synthesis of Flexible, Ultrathin Gold Nanowires in
  Organic Media}. Langmuir 24~(17), 9855--9860.
\newline\urlprefix\url{http://pubs.acs.org/doi/abs/10.1021/la801675d}

\bibitem[{Plateau(1873)}]{Plateau1873}
Plateau, J. A.~F., 1873. {Statique experimentale et theorique des liquides
  soumis aux seules forces moleculaires}. Gauthiers-Villars, Paris.

\bibitem[{Rayleigh(1878)}]{Rayleigh1878}
Rayleigh, J. W.~S., 1878. {On the Instability of Jets}. Proc. London Math. Soc.
  10~(4).

\bibitem[{Reiser et~al.(2016)Reiser, Gerstner, Gonzalez-Garcia, Maurer,
  Kanelidis, and Kraus}]{Reiser2016}
Reiser, B., Gerstner, D., Gonzalez-Garcia, L., Maurer, J. H.~M., Kanelidis, I.,
  Kraus, T., 2016. {Multivalent bonds in self-assembled bundles of ultrathin
  gold nanowires}. Physical Chemistry Chemical Physics 18~(39), 27165--27169.
\newline\urlprefix\url{http://xlink.rsc.org/?DOI=C6CP05181B}

\bibitem[{Reiser et~al.(2017)Reiser, Gerstner, Gonzalez-Garcia, Maurer,
  Kanelidis, and Kraus}]{Reiser2017}
Reiser, B., Gerstner, D., Gonzalez-Garcia, L., Maurer, J. H.~M., Kanelidis, I.,
  Kraus, T., may 2017. {Spinning Hierarchical Gold Nanowire Microfibers by
  Shear Alignment and Intermolecular Self-Assembly}. ACS Nano 11~(5),
  4934--4942.
\newline\urlprefix\url{http://pubs.acs.org/doi/10.1021/acsnano.7b01551}

\bibitem[{Repko and Cademartiri(2012)}]{Repko2012}
Repko, A., Cademartiri, L., dec 2012. {Recent advances in the synthesis of
  colloidal nanowires}. Canadian Journal of Chemistry 90~(12), 1032--1047.
\newline\urlprefix\url{http://www.nrcresearchpress.com/doi/10.1139/v2012-077}

\bibitem[{Sadasivan et~al.(2005)Sadasivan, Khushalani, and
  Mann}]{Sadasivan2005}
Sadasivan, S., Khushalani, D., Mann, S., may 2005. {Synthesis of Calcium
  Phosphate Nanofilaments in Reverse Micelles}. Chemistry of Materials 17~(10),
  2765--2770.
\newline\urlprefix\url{http://pubs.acs.org/doi/abs/10.1021/cm047926g}

\bibitem[{S{\'{a}}nchez-Iglesias et~al.(2012)S{\'{a}}nchez-Iglesias,
  Rivas-Murias, Grzelczak, P{\'{e}}rez-Juste, Liz-Marz{\'{a}}n, Rivadulla, and
  Correa-Duarte}]{SanchezIglesias2012}
S{\'{a}}nchez-Iglesias, A., Rivas-Murias, B., Grzelczak, M., P{\'{e}}rez-Juste,
  J., Liz-Marz{\'{a}}n, L.~M., Rivadulla, F., Correa-Duarte, M.~A., dec 2012.
  {Highly Transparent and Conductive Films of Densely Aligned Ultrathin Au
  Nanowire Monolayers}. Nano Letters 12~(12), 6066--6070.
\newline\urlprefix\url{http://pubs.acs.org/doi/10.1021/nl3021522}

\bibitem[{Shin et~al.(2007)Shin, Yu, and Song}]{Shin2007}
Shin, H.~S., Yu, J., Song, J.~Y., oct 2007. {Size-dependent thermal instability
  and melting behavior of Sn nanowires}. Applied Physics Letters 91~(17),
  173106.
\newline\urlprefix\url{http://aip.scitation.org/doi/10.1063/1.2801520}

\bibitem[{Takahata et~al.(2016)Takahata, Yamazoe, Warakulwit, Limtrakul, and
  Tsukuda}]{Takahata2016}
Takahata, R., Yamazoe, S., Warakulwit, C., Limtrakul, J., Tsukuda, T., aug
  2016. {Rayleigh Instability and Surfactant-Mediated Stabilization of
  Ultrathin Gold Nanorods}. The Journal of Physical Chemistry C 120~(30),
  17006--17010.
\newline\urlprefix\url{http://pubs.acs.org/doi/10.1021/acs.jpcc.6b03113}

\bibitem[{{Toimil Molares} et~al.(2004){Toimil Molares}, Balogh, Cornelius,
  Neumann, and Trautmann}]{Toimil-Molares2004}
{Toimil Molares}, M.~E., Balogh, A.~G., Cornelius, T.~W., Neumann, R.,
  Trautmann, C., nov 2004. {Fragmentation of nanowires driven by Rayleigh
  instability}. Applied Physics Letters 85~(22), 5337--5339.
\newline\urlprefix\url{http://aip.scitation.org/doi/10.1063/1.1826237}

\bibitem[{Wang et~al.(2007)Wang, Hou, Kim, and Sun}]{Wang2007}
Wang, C., Hou, Y., Kim, J., Sun, S., aug 2007. {A General Strategy for
  Synthesizing FePt Nanowires and Nanorods}. Angewandte Chemie International
  Edition 46~(33), 6333--6335.
\newline\urlprefix\url{http://doi.wiley.com/10.1002/anie.200702001}

\bibitem[{Wang et~al.(2008)Wang, Hu, Lieber, and Sun}]{Wang2008}
Wang, C., Hu, Y., Lieber, C.~M., Sun, S., jul 2008. {Ultrathin Au Nanowires and
  Their Transport Properties}. Journal of the American Chemical Society
  130~(28), 8902--8903.
\newline\urlprefix\url{http://pubs.acs.org/doi/abs/10.1021/ja803408f}

\bibitem[{Wu et~al.(2015)Wu, Pan, Su, and Yang}]{WuJianbo2015}
Wu, J., Pan, Y.-T., Su, D., Yang, H., aug 2015. {Ultrathin and stable AgAu
  alloy nanowires}. Science China Materials 58~(8), 595--602.
\newline\urlprefix\url{http://link.springer.com/10.1007/s40843-015-0072-z}

\bibitem[{Xi et~al.(2006)Xi, Liu, Wang, Liu, Peng, and Qian}]{Xi2006}
Xi, G., Liu, Y., Wang, X., Liu, X., Peng, Y., Qian, Y., nov 2006. {Large-Scale
  Synthesis, Growth Mechanism, and Photoluminescence of Ultrathin Te
  Nanowires}. Crystal Growth {\&} Design 6~(11), 2567--2570.
\newline\urlprefix\url{http://pubs.acs.org/doi/abs/10.1021/cg0603218}

\bibitem[{Xu et~al.(2013)Xu, Zhu, Zhu, and Jiang}]{Xu2013}
Xu, J., Zhu, Y., Zhu, J., Jiang, W., 2013. {Ultralong gold nanoparticle/block
  copolymer hybrid cylindrical micelles: a strategy combining surface templated
  self-assembly and Rayleigh instability}. Nanoscale 5~(14), 6344.
\newline\urlprefix\url{http://xlink.rsc.org/?DOI=c3nr01296d}

\bibitem[{Xu et~al.(2018)Xu, Li, and Lu}]{Xu2017}
Xu, S., Li, P., Lu, Y., feb 2018. {In situ atomic-scale analysis of Rayleigh
  instability in ultrathin gold nanowires}. Nano Research 11~(2), 625--632.
\newline\urlprefix\url{http://link.springer.com/10.1007/s12274-017-1667-3}

\bibitem[{Yu et~al.(2006)Yu, Joo, Park, and Hyeon}]{Yu2006}
Yu, T., Joo, J., Park, Y.~I., Hyeon, T., feb 2006. {Single Unit Cell Thick
  Samaria Nanowires and Nanoplates}. Journal of the American Chemical Society
  128~(6), 1786--1787.
\newline\urlprefix\url{http://pubs.acs.org/doi/abs/10.1021/ja057264b}

\bibitem[{Zhao et~al.(2006)Zhao, Averback, and Cahill}]{Zhao2006}
Zhao, K., Averback, R.~S., Cahill, D.~G., jul 2006. {Patterning of metal
  nanowires by directed ion-induced dewetting}. Applied Physics Letters 89~(5),
  053103.
\newline\urlprefix\url{http://aip.scitation.org/doi/10.1063/1.2261271}

\bibitem[{Zhao et~al.(2016)Zhao, Huang, Yuan, and Wang}]{Zhao2016}
Zhao, W., Huang, D., Yuan, Q., Wang, X., oct 2016. {Sub-2.0-nm Ru and
  composition-tunable RuPt nanowire networks}. Nano Research 9~(10),
  3066--3074.
\newline\urlprefix\url{http://link.springer.com/10.1007/s12274-016-1189-4}

\bibitem[{Zhou et~al.(2009)Zhou, Zhou, Pan, Lei, and Xu}]{Zhou2009}
Zhou, Z., Zhou, Y., Pan, Y., Lei, W., Xu, C., apr 2009. {Overheating and
  undercooling of Ni polycrystalline nanowires}. Scripta Materialia 60~(7),
  512--515.
\newline\urlprefix\url{http://linkinghub.elsevier.com/retrieve/pii/S1359646208008397}

\end{thebibliography}

\end{document}